\documentclass[letterpaper]{article} 
\usepackage{aaai25}  
\usepackage{times}  
\usepackage{helvet}  
\usepackage{courier}  
\usepackage[hyphens]{url}  
\usepackage{graphicx} 
\usepackage{natbib}  
\usepackage{caption} 
\usepackage{algorithm}
\usepackage{algorithmic}

\usepackage{amsmath}
\usepackage{booktabs}
\usepackage{array}

\usepackage{xcolor}
\newcommand{\answerYes}[1]{\textcolor{blue}{#1}} 
\newcommand{\answerNo}[1]{\textcolor{teal}{#1}} 
\newcommand{\answerNA}[1]{\textcolor{gray}{#1}} 
\newcommand{\answerTODO}[1]{\textcolor{red}{#1}} 
\usepackage{newfloat}
\usepackage{listings}
\DeclareCaptionStyle{ruled}{labelfont=normalfont,labelsep=colon,strut=off} 
\floatstyle{ruled}
\newfloat{listing}{tb}{lst}{}
\floatname{listing}{Listing}
\title{Algorithmic Behaviors Across Regions: A Geolocation Audit of YouTube Search for COVID-19 Misinformation Between the United States and South Africa}

\author {
    Hayoung Jung\textsuperscript{\rm 1},
    Prerna Juneja\textsuperscript{\rm 2},
    Tanushree Mitra\textsuperscript{\rm 1}
}
\affiliations {
    \textsuperscript{\rm 1}University of Washington\\
    \textsuperscript{\rm 2}Seattle University\\
    hjung10@uw.edu, pjuneja@seattleu.edu, tmitra@uw.edu
}

\begin{document}

\maketitle
\begin{abstract}
Despite being an integral tool for finding health-related information online, YouTube has faced criticism for disseminating COVID-19 misinformation globally to its users. Yet, prior audit studies have predominantly investigated YouTube within the Global North contexts, often overlooking the Global South. To address this gap, we conducted a comprehensive 10-day geolocation-based audit on YouTube to compare the prevalence of COVID-19 misinformation in search results between the United States (US) and South Africa (SA), the countries heavily affected by the pandemic in the Global North and the Global South, respectively. For each country, we selected 3 geolocations and placed sock-puppets, or bots emulating ``real'' users, that collected search results for 48 search queries sorted by 4 search filters for 10 days, yielding a dataset of 915K results. We found that 31.55\% of the top-10 search results contained COVID-19 misinformation. Among the top-10 search results, bots in SA faced significantly more misinformative search results than their US counterparts. Overall, our study highlights the contrasting algorithmic behaviors of YouTube search between two countries, underscoring the need for the platform to regulate algorithmic behavior consistently across different regions of the Globe.

\answerTODO{\textit{Warning: We caution the readers that some examples provided to better contextualize our data can be offensive.}}

\end{abstract}

\begin{figure}[!ht]
  \small
  \centering
  \includegraphics[width=0.6275\linewidth]{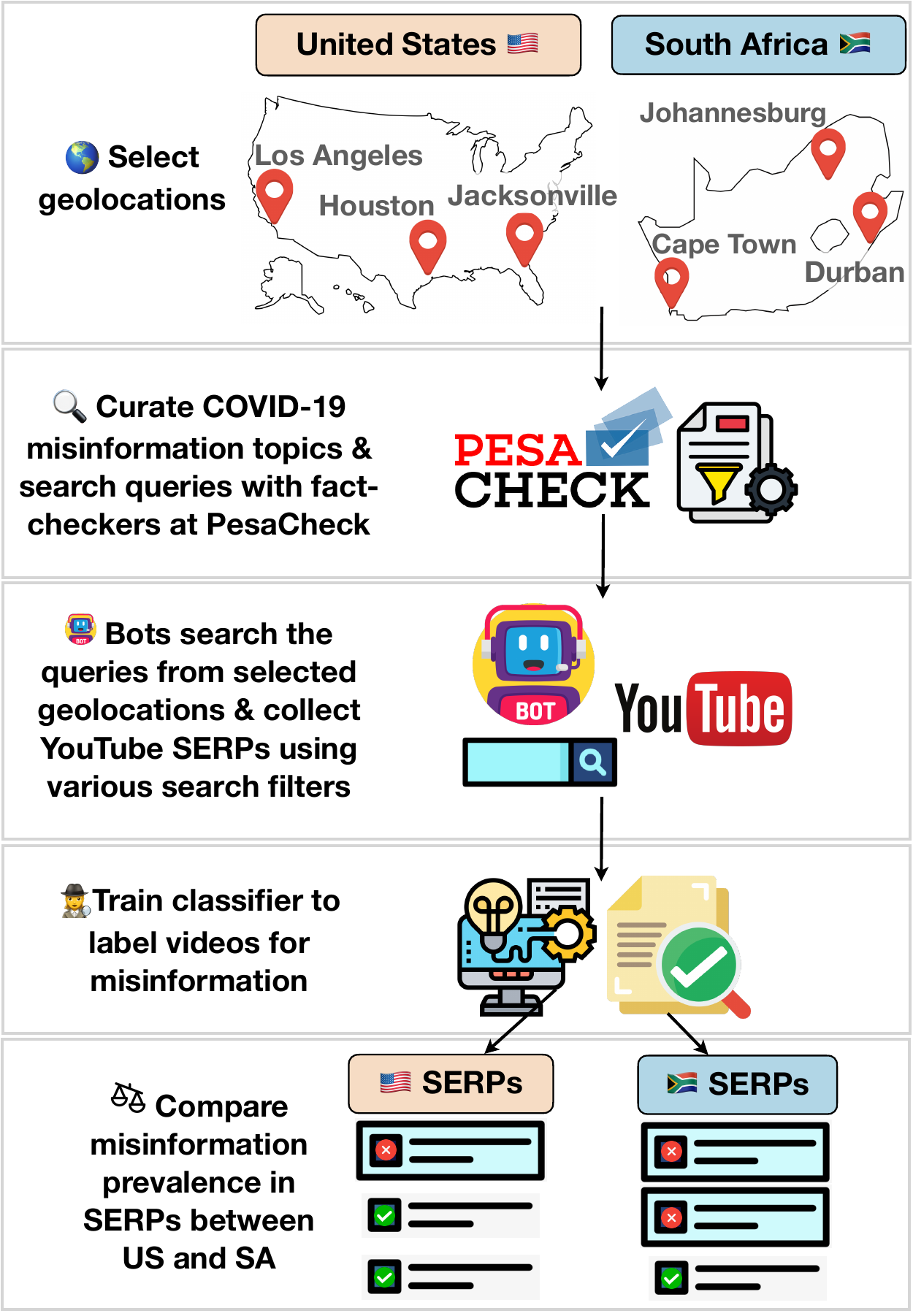}
  \caption{\textbf{Pipeline Overview.} Sock-puppet bots emulating real-world users utilized the curated search queries to gather YouTube search engine result pages (SERPs) from geolocations in the United States (US) and South Africa (SA). After training and employing a classifier to scale the video labeling process, we compared the prevalence of COVID-19 misinformation in SERPs between the two countries.}
  \label{map}
\end{figure}

\section{Introduction}

\begin{quote}
    \small
    \textit{``This virus is here to stay. It is still killing and it is still changing.''}
    --- \citet{still_pandemic2}, Director-General of the World Health Organization
\end{quote}

Since March 2020, the World Health Organization (WHO) has designated COVID-19 as a global pandemic. The pandemic continues to pose a public health threat, registering 206K COVID-19 cases and 3.5K deaths worldwide in July 2024 alone \cite{WHO}. 
As the largest video search engine, YouTube has emerged as a vital tool for finding health-related information online, particularly during outbreaks and global pandemics \cite{dubey2014analysis, bora2018internet, KHATRI2020101636}.
However, YouTube has faced criticism for disseminating COVID-19 misinformation globally to its users, with fact-checkers calling the platform a ``major conduit of fake news'' \cite{factchecker}. This misinformation has undermined public health efforts worldwide, fueling vaccine hesitancy and eroding trust in health institutions \cite{who_infodemic}.
Thus, there is a pressing need for empirical investigation of search engine systems to ensure algorithmic accountability, safeguard global health, and promote a more responsible and trustworthy web. 

In response to public pressure, YouTube collaborated with the WHO to develop a content moderation policy for COVID-19 misinformation for their platform \cite{google_misinfo}. However, past reports expressed concerns that YouTube's content moderation practices are biased in favor of the Global North and neglect the misinformation challenges in the Global South, noting that YouTube does not fact-check or remove videos in non-English languages \cite{facebook_youtube, nyt_youtube}. While multiple previous studies have empirically investigated search engines for COVID-19 misinformation in the Global North context \cite{papadamou, doi:10.1177/14648849231157845}, the Global South has received little attention. The global aspect of the pandemic presents a unique and crucial opportunity to conduct a comparative audit of COVID-19 misinformation in two different parts of the world.

Guided by an overarching research question: \textit{What is the prevalence of COVID-19 misinformation in YouTube search results between the US and SA?}, we conducted a comprehensive 10-day geolocation-based comparative audit of YouTube search from January 30th, 2023 to February 9th, 2023. The goal of the audit was to compare the prevalence of COVID-19 misinformation in Search Engine Result Pages (SERPs) between the United States (US) and South Africa (SA), the countries heavily affected by the pandemic in the Global North and the Global South, respectively. For each country, we selected 3 geolocations and placed sock-puppets (bots emulating real users) that collected YouTube SERPs for 48 search queries belonging to 8 globally persistent COVID-19 misinformation topics, such as ``Bill Gates Claims.'' To gain deeper insights into the platform's sorting algorithm, we sorted the search results across 4 search filters: the default ``Relevance,'' ``Upload Date,'' ``View Count,'' and ``Rating,'' resulting in 915K search results. We scored the videos based on their stance toward COVID-19 misinformation and compared the amount of misinformation present in search results between the US and SA. 

We find multiple instances where bots in SA encountered statistically significantly more misinformative SERPs than bots in the US. These disparities were observed within the top-10 search results (p$<$0.001, \textit{r}=0.49) and search results sorted by YouTube's ``Relevance'' filter (p$<$0.001, \textit{r}=0.86) with medium to large effect sizes, indicating practical significance. Since 95\% of the user traffic is directed towards the first page of the search results \cite{web_traffic} and YouTube employs the ``Relevance'' filter by default, users in SA may likely encounter significantly more misinformative SERPs than users in the US.
Given YouTube's established importance for finding health information, this could raise personal health risks for users in SA by potentially negatively influencing their beliefs, health practices, and decisions. 
Overall, our work highlights the contrasting algorithmic behaviors of YouTube's search function in two countries within the context of COVID-19 misinformation, underscoring the need for YouTube to regulate its algorithmic behaviors consistently across different regions of the Globe. 

\subsubsection{Contributions and Implications.} To our knowledge, our study is the first large-scale geolocation-based comparative audit of YouTube search for COVID-19 misinformation across countries in the Global North and the Global South. This research explores the less-explored context of the Global South, extending the scope of previous algorithmic audits, which primarily centered on the Global North. Our work provides an elaborate understanding of YouTube's search engine across 4 distinct search filters for 8 misinformation topics. Through a 10-day data collection and extensive labeling process, we amassed a labeled dataset\footnote{\url{https://github.com/social-comp/YouTubeAuditGeolocation-data}} of 915K search results (10,139 unique videos) and trained a classifier\footnote{https://huggingface.co/SocialCompUW/youtube-covid-misinfo-detect} to detect COVID-19 misinformation in YouTube videos. 
Overall, our work highlights the contrasting algorithmic behaviors of YouTube's search function in two countries within the context of COVID-19 misinformation, underscoring the need for YouTube to regulate its algorithmic behaviors consistently across different regions of the Globe.

\section{Related Work}

\subsection{Algorithmic Audits of Search Engines}
Search engines determine what information is relevant and shape user behavior, impacting aspects such as political voting behavior \cite{diakopoulos2018vote}, scientific knowledge \cite{papadamou}, and beliefs \cite{doi:10.1177/1075547015596367}. Despite their societal importance, search engines operate without external regulation, leaving the credibility of content unverified. Consequently, researchers have investigated search systems by conducting algorithmic audits that empirically measure and understand the conditions under which problematic content arises on the platform. Several studies have investigated search engines for misinformation \cite{10.1145/3392854, papadamou, 10.1145/3460231.3474241}, conspiracy theories \cite{doi:10.1126/sciadv.add8080, faddoul2020longitudinal}, hate speech and extremism \cite{10.1145/3555618, 10.1145/3351095.3372879}, and partisanship \cite{10.1145/3274417}. 
Among the various methods for auditing search engines, we utilize sock-puppet audits (programming bots to emulate real users), a methodology commonly employed in prior audits \cite{10.1145/3392854,  papadamou, 10.1145/2815675.2815714} for its control over experimental variables. Our study adds to the existing sock-puppet audit studies, investigating the prevalence of COVID-19 misinformation on YouTube between the US and SA.

\subsection{(Lack of) Algorithmic Audits in the Global South} 

Prior works have highlighted the growing need to consider the Global South in algorithmic audit research \cite{10.1145/3531146.3533213}. In a recent paper, \citet{Urman2024WEIRDAR} argued that most algorithmic audit research is skewed towards the Western context, noting that ``countries located outside of North America and Western Europe are understudied.'' A limited number of studies have conducted algorithmic audits within the Global South context, examining factors such as language \cite{narain2023covid} and culture \cite{dammu-etal-2024-uncultured}. Among these audits conducted in the Global South contexts, only a few considered geolocation, with a vast majority focusing on Google Search \cite{le2022crowdsourcing, dabran2023covid}. We contribute to the less-researched context of the Global South and extend prior algorithmic audits on YouTube, the largest video search engine.

\subsection{Search-Enabled COVID-19 Misinformation} 

Given the importance of search engines for finding health-related information during the COVID-19 pandemic \cite{KHATRI2020101636}, several scholars have audited search engines for COVID-19 misinformation \cite{Lie002604, doi:10.1177/14648849231157845, papadamou} and anti-vaccine content \cite{10.1145/3411764.3445250, papadamou}. In addition, researchers have engineered various features and built machine learning models to automatically detect COVID-19 misinformation online \cite{info:doi/10.2196/49061, papadamou, medina-serrano-etal-2020-nlp}. However, most prior studies have primarily focused on the Global North, leaving a gap in audits of search engines for COVID-19 misinformation in the Global South contexts. 

Existing literature regarding the Global South has several gaps, often probing search engines with the default search filter for a single day and focusing on a narrow range of topics. For example, researchers in \citet{narain2023covid} focused on the language aspect of the audit, examining COVID-19 videos in 11 widely spoken languages across Africa on YouTube and collecting 562 videos for evaluation. \citet{dabran2023covid} examined geolocation and language, analyzing 3 COVID-19 conspiracy topics on Google Search across 4 languages in 10 countries for a single day, collecting 330 search results. We systematically compare the prevalence of COVID-19 misinformation on YouTube across countries in the Global North and Global South. Across geolocations in the US and SA, we examine 8 different COVID-19 misinformation topics and 4 distinct search filters over 10 days of data collection, yielding 915K search results. Unlike previous audits with single geolocations per country, we selected three per country for fine-grained comparative analysis within countries.  

\section{Audit Experiment Setup}\label{sec:methods}
This section presents the methodology for selecting geolocations for our audit experiments, curating globally persistent COVID-19 misinformation topics and associated search queries, and designing our experimental setup.

\subsection{Selecting the Geolocations for the Audit}\label{sec:geolocation}

We considered the US and SA because they were heavily affected by the pandemic in the Global North and the Global South, respectively, making them vulnerable to COVID-19 misinformation\footnote{As of May 7th, 2024, the US and SA continue to have the highest reported cases of COVID-19 in the continents of North America and Africa, respectively \cite{WHO}.} \cite{WHO}. Previous work established that Google personalizes search results across different states in the US \cite{10.1145/2815675.2815714}. Therefore, we chose to identify three states in each country to achieve a more robust analysis of COVID-19 misinformation in YouTube search results of the selected countries. We identified three states in the US and three provinces\footnote{South Africa operates on a provincial government system.} in South Africa with the highest total confirmed cases of COVID-19 \cite{npr_us_cases, sa_covid2}, which may make them susceptible to COVID-19 misinformation. 
To capture the highest proportion of the population in these states and provinces, we selected the largest populated city as the geolocation. For US, we selected Los Angeles (California), Houston (Texas), and Jacksonville (Florida). For South Africa, we selected Johannesburg (Gauteng), Durban (KwaZulu-Natal), and Cape Town (Western Cape). Figure \ref{map} depicts the geolocations chosen in the US and SA. 

\begin{table}[!t]
\centering
\scriptsize
\begin{tabular}{ll}
\toprule
\textbf{Misinformation Topics} & \textbf{Sample Search Queries} \\ \hline
Biological Weapon & 
  Biological Weapon, man-made virus \\\cline{2-2}
Lab Leak Theory & 
  lab leak theory, Kungflu \\\cline{2-2}
5G Claims & 
  5g and covid19 link, 5g conspiracy \\\cline{2-2}
Bill Gates Claims & 
  bill gates exposed, bill gates vaccine chip \\\cline{2-2}
Spread of Virus & 
  social spread, sanitize \\\cline{2-2}
Treatment of Virus & 
  local concoctions, sesame oil \\\cline{2-2}
Population Control & 
  population control, plandemic \\\cline{2-2}
Vaccine Content Claims & 
  MRNA, fetal tissue research \\
\bottomrule
\end{tabular}
\caption{The 8 globally-persistent COVID-19 misinformation topics identified by \textit{PesaCheck}. For each topic, we provide a sample of our curated search queries.}
\label{tab:topics-query-sample}
\end{table}

\subsection{Curating Topics and Search Queries}\label{sec:compile}
\subsubsection{Curating COVID-19 Misinformation Topics.}
To curate globally persistent COVID-19 misinformation topics, we partnered with expert fact-checkers at \textit{Pesacheck}.\footnote{https://www.pesacheck.org} As Africa's largest indigenous fact-checking organization, \textit{PesaCheck} is affiliated with the International Fact-Checking Network (IFCN) and collaborates with expert fact checkers worldwide, making them well equipped to identify misinformation that circulates in diverse regions worldwide. 

Using editorial coverage, frequently fact-checked information and social media analytics tools, the expert fact-checkers identified eight globally persistent COVID-19 misinformation topics and provided a dataset of 362 fact-checked YouTube videos related to these topics. These videos originate from channels associated with 29 countries, spanning the Global North and the Global South, including Algeria, China, South Africa, and the United States. This diverse representation--18 countries from the Global North and 11 from the Global South (see Appendix Table 3)--ensures that the search queries developed from these videos are representative across different geographic regions. Refer to Appendix Table \ref{tab:topics-sample} for sample YouTube videos provided by \textit{PesaCheck}. Table \ref{tab:topics-query-sample} presents the 8 topics and samples of our curated search queries, which we explain next. 

\subsubsection{Curating Search Queries.}\label{curation}
We utilized three methods to curate search queries for each topic. Given that English is the only language spoken commonly in both the US and SA,\footnote{English is one of eleven official languages in SA, with 10\% of people speaking English natively at home and is most commonly spoken in business and commerce \cite{SA_language_statistics}.} we focused on queries in English, allowing for a controlled cross-country comparison in the amount of COVID-19 misinformation returned by YouTube's search algorithm.

First, we used YouTube video tags in the videos provided by \textit{Pesacheck}. Video tags are descriptive keywords that represent how content creators want their videos to be discovered \cite{youtube_tags}. Misinformative videos often contain tags that describe misleading narratives, effectively serving as potential search queries to identify more misinformative videos \cite{10.1145/3544548.3580846}.
We collected 2,911 video tags from the videos and applied systematic filtering criteria used in previous work \cite{10.1145/3411764.3445250}. We excluded queries mentioning individuals or news organizations (e.g., ``Obama'', ``Republic TV''), overly generic terms (e.g., ``breaking news''), excessively specific terms (e.g., ``COVID3rdWaveInMyanmar''), and irrelevant terms (e.g., ``cute puppies''). Additionally, we manually merged similar queries (e.g. ``man-made'' and ``man-made virus'' into the single query ``man-made virus''). These systematic filtering steps resulted in 48 tags. For example, the tags included problematic terms regarding the origin of the virus in China (``Kungflu,'' ``China virus'') and terms on self-treatment methods (``local concoction'').

Second, for each topic, we identified important keywords found in video titles, descriptions, and transcripts from \textit{PesaCheck}-provided videos. We created a document for each topic by concatenating the video metadata and applying standard preprocessing steps. To extract keywords from each topic document, we used \textit{KeyBERT} \cite{keybert}, a keyword extraction tool that requires two inputs: an embedding model and a document. First, \textit{KeyBERT} applies the embedding model to the topic document to generate the document embeddings and extract word embeddings for N-gram phrases.\footnote{We limited our phrases to unigram, bigram, and trigrams.} Then, \textit{KeyBERT} uses cosine similarity to identify keywords with the highest similarity with the document itself, providing words that best describe the entire document. We chose the Sentence Transformer \textit{all-mpnet-base-v2} due to their best performance in generating sentence embeddings \cite{song2020mpnet}. Since the model was not trained on COVID-19-related texts, we fine-tuned the model for domain adaptation using COVID-19 question-answering datasets \cite{oller2020covidqa, covid-qa, tang2020rapidly}. See Appendix \ref{sec:finetuning} for fine-tuning details. After fine-tuning, we used \textit{KeyBERT} on each of the 8 topic documents to extract 5 keywords per document, resulting in 40 search queries.

Third, we included the 49 search keywords provided by \textit{PesaCheck} because the keywords were curated by expert fact-checkers to find misinformative videos. Combining the search queries from all three methods, we obtained 137 search queries. Then, we manually removed duplicate queries, combined similar queries, and randomly selected 6 search queries per topic, resulting in a final set of 48 search queries (see Appendix Table \ref{tab:full-query-set}). To ensure that the search queries output search results relevant to the COVID-19 pandemic, we identified keyword variations of COVID-19 and employed YouTube search operators, which are special commands that can be utilized to efficiently refine their searches on YouTube \cite{search_operator}. We selected 8 keyword variations of COVID-19 using a COVID-19 Twitter dataset \cite{info:doi/10.2196/19273}. 
Each query was formatted as:

\noindent\texttt{\small\textbf{query} (covid {\textbar} corona {\textbar}  covid-19 {\textbar}  covid19 {\textbar} coronavirus {\textbar} COVD {\textbar} sars-cov-2 {\textbar} pandemic)}

\subsection{Experimental Design}\label{sec:design}
\subsubsection{Overview.}\label{sec:audit_overview}
To host our experiments, we used Amazon Web Services (AWS) to create all the virtual machines (VMs). We programmed Selenium bots \cite{selenium} to emulate real-world users and automate browser actions. To obscure the automated interactions of the bots, we followed the suggestions from \citet{hide_bot}. Each bot utilized \textit{IPRoyal} proxies \cite{iproyal} and validated the IP geolocation of the proxy using \textit{IP2Location}, an IP geolocation lookup service \cite{ip2location}, to obtain personalized search results from the desired IP geolocation. For each query, we collected YouTube SERPs sorted by 4 search filters: ``Relevance,'' ``Upload Date,'' ``View Count,'' and ``Rating.'' During the data collection, we extracted the top 50 search results from each SERP. Additionally, we added wait times after every browser action and chose two evenly separate times to distribute our search queries throughout the day to avoid getting rate-limited by YouTube. We ran the audit experiment for 10 consecutive days from January 30th, 2023 to February 9th, 2023, where we simultaneously searched 24 queries at 00:00 UTC and the other 24 queries at 12:00 UTC. 

To control for possible confounding factors that may affect our audit, we followed standard noise control procedures based on prior work \cite{10.1145/2815675.2815714, 10.1145/3392854, 10.1145/3411764.3445250}. 
To differentiate between noise and geolocation-based personalization in SERPs, we created identical twin bots, consisting of a treatment bot and its corresponding control bot that conducted the same actions simultaneously at each geolocation. Thus, any difference in the search results between the twin bots should be attributed to noise rather than personalization.\footnote{Note that YouTube’s search engine is a black box. Even after controlling for all known sources of noise, there could be some sources of noise we are unaware of.} If the differences in the SERPs between two geolocations exceeded the noise, it can only be attributed to geolocation-based personalization. In our experiment, we placed twin bots at each geolocation, resulting in 12 bots. 

\subsubsection{Validation Experiments.} 

Changing IP addresses using VMs \cite{10.1145/3392854} and proxies \cite{10.1145/3485447.3512102} are common methods to conduct geolocation audits, especially in the US context. However, obtaining \textit{fine-grained} coverage of IP addresses through VMs and proxy services in South Africa was challenging. The only AWS coverage in South Africa, let alone in Africa, is in Cape Town. Meanwhile, nearly all the proxy and VPN providers except \textit{IPRoyal} were unfeasible due to the steep pricing or the lack of proxy coverage of our selected geolocations in South Africa (see Appendix Table \ref{tab:proxy} for further details). Due to challenges associated with IP addresses, we initially turned to the ``geospoofing'' method used in \citet{10.1145/2815675.2815714}, in which they fed precise latitude and longitude coordinate information to automated scripts and obtained personalized search results, providing a cost-effective alternative. For brevity, we summarized the results from our validation experiments and left the details in the Appendix. We defined the metric for geolocation-based personalization in Appendix \ref{sec:personalization_metric}. In our first validation experiment, we curated a set of search queries and validated that the queries elicited geolocation-based personalization, resulting in personalized SERPs based on geolocation (Appendix \ref{sec:appendix_val1}). Using the validated search queries, we performed a second validation experiment and found that YouTube uses IP geolocation instead of geospoofed location to personalize search results (Appendix \ref{sec:appendix_val2}). Thus, we conducted a third validation experiment using \textit{IPRoyal} proxies, which validated the accuracy and consistency of the proxies in giving us the correct IP geolocations for our experiment (Appendix \ref{sec:appendix_val3}).

\section{Developing Data Annotation Scheme}\label{sec:annotation}
Our geolocation audit experiment collected 23,020 SERPs\footnote{The 12 bots scraped 48 queries across 4 search filters for 10 days, resulting in $48 \times 4 \times 10 \times 12 = 23,040$ SERPs. On the 8th day of the experimental run, a sock-puppet bot in Durban crashed due to technical errors, failing to collect 20 SERPs. We excluded the queries from this particular experimental run from the analysis.} consisting of 915,440 search results (10,139 unique videos). To label these videos for COVID-19 misinformation, we underwent extensive procedures to determine what constitutes misinformation and develop the qualitative coding scheme. 

\subsection{How Do We Know What Is Misinformation?}

To determine what constitutes misinformation, we based our annotation heuristics on Google’s COVID-19 medical misinformation policy, which has been developed in partnership with the WHO \cite{google_misinfo}. We also referenced the policies of national health authorities in the US and SA, such as the Center for Disease Control and Prevention (CDC) \cite{cdc} and the South African Government’s COVID-19 Fake News resources \cite{sa_fakenews}. Given the misleading nature of xenophobic terms such as ``Kungflu'' and ``China Virus,'' we also incorporated them in our annotation heuristics. Additionally, we took extra precautions in annotating videos about the COVID-19 Lab Leak Theory\footnote{The Lab Leak Theory is contentious, arguing that COVID-19 leaked from a lab in Wuhan, China.} by referencing a declassified report from the US National Intelligence Council, which presented the consensus by various government agencies \cite{intelligence}. The report assessed that both the Natural Origins Theory\footnote{The Natural Origins Theory contests that COVID-19 spread to humans from animals, such as bats and pangolins.} and the Lab Leak Theory are plausible until more evidence comes to light. However, the report also debunked many misleading claims, such as COVID-19 being a biological weapon developed in a lab. Thus, we did not label videos as misinformative for discussing the origin theories about the virus (such videos were labeled as ``On the COVID-19 origins in Wuhan, China'' -- see Appendix Table \ref{tab:annotation}); however, we labeled videos as misinformative if they promoted debunked and misleading claims outlined in the report.

\subsection{Annotation Scale and Heuristics}\label{annotation_dev}
Developing the qualitative coding scheme to label YouTube videos for COVID-19 misinformation was challenging, requiring multiple iterations and discussions with external researchers to refine the heuristics. In the first iteration, the first author sampled 80 videos from the audit experiment and annotated the videos. After multiple iterations analyzing each video, the author created an initial 7-point annotation scale: ``Opposing COVID-19 Misinformation (-1),'' ``Neutral COVID-19 Information (0),'' ``Supporting COVID-19 Misinformation (1)'', ``On the COVID-19 origins in Wuhan, China (2),'' ``Irrelevant (3),'' ``Video in a language other than English (4),'' and ``URL not accessible (5).'' Next, seven external researchers with extensive work experience in online misinformation independently annotated 13 videos and provided feedback on our annotation criteria and tasks. After discussion and incorporating their feedback, we further refined the annotation heuristics. Due to space constraints, please see Appendix Table \ref{tab:annotation} for the 7-point annotation labels, descriptions, and example videos. 

\section{Labeling YouTube Videos}\label{sec:classifier}
After developing our data annotation scheme, we labeled the videos.  Given the large amounts of data (10,139 videos), we scaled the labeling process using a machine-learning classifier for English videos. We constructed the ground-truth dataset, trained and evaluated 62 different classifiers, and separately handled videos in non-English languages.

\subsection{Creating the Ground-Truth Dataset}\label{ground-truth}

We obtained annotations for 3,075 videos, which were annotated by the first author and Amazon Mechanical Turk (AMT). The first author, as the expert, annotated 1,087 videos.\footnote{To gather more annotation, the first author followed \citet{10.1145/3460231.3474241} and reannotated \citet{medina-serrano-etal-2020-nlp}'s YouTube video dataset, resulting in 143 more annotations.} For 1,988 videos, we obtained three AMT worker annotations per video. To assess the AMT workers' agreement, we calculated the Fleiss' Kappa Score and obtained $\kappa=0.62$, indicating ``substantial agreement.'' 
We employed the majority response to assign the final label, arriving at a final label for 1,899 videos. For the remaining 89 videos, all three AMT worker responses diverged. The first author annotated the 89 videos to obtain the final annotation values. In \S\ref{sec:ethics}, we address the steps to minimize potential harm associated with exposing AMT workers to misinformation. See Appendix \ref{sec:mturk} for the AMT worker training, screening, compensation, and annotation task. 
Overall, the ground-truth dataset consisted of 3,075 videos, 820 of which were supporting, 837 opposing, 431 neutral, 409 irrelevant, 317 non-English, 228 on COVID-19 origins, and 33 URLs not accessible. To train our classifier, we excluded videos annotated as non-English and URL not accessible.

\subsection{Training and Applying Classifier to English Videos}\label{sec:classifer_training}

Using the ground-truth dataset, we trained 62 different classifiers to find the best-performing model for our task. 

\subsubsection{Consolidating From 5-Classes to 3-Classes.} Developing a classifier to detect COVID-19 misinformation in YouTube videos was difficult, requiring experimentation with several models. Initially, we trained classifiers that predicted five classes: opposing, supporting, neutral, COVID-19 origins, and irrelevant. Such a model could only achieve an accuracy of 0.71. We sought to improve our classifier's accuracy by reducing the number of classes, while maintaining our study's objective of measuring the level of COVID-19 misinformation in YouTube SERPs. Thus, we merged the classes neutral, irrelevant, and COVID-19 origins into a single category as these classes \textit{neither} support nor oppose COVID-19 misinformation. This consolidation yielded a classification task with 3-classes: supporting misinformation, opposing misinformation, and neither.

\begin{table}[!t]
\centering
\small
\begin{tabular}{l|ccc}  
\toprule
\textbf{Model} & \textbf{Acc.} & \textbf{F1-M} & \textbf{F1-W} \\ \midrule
SVM & 0.78 & 0.78 & 0.78 \\
XGB & 0.75 & 0.75 & 0.76 \\
DeBerta (base) & 0.81 & 0.81 & 0.81 \\
DeBerta (large) & \textbf{0.85} & \textbf{0.85} & \textbf{0.85} \\
GPT-4 Turbo & 0.79 & 0.79 & 0.79 \\
\bottomrule
\end{tabular}
\caption{The best performance achieved by each model on the three-class classification task: Support Vector Machine (SVM), XGBoost (XGB), DeBerta-v3-base, DeBerta-v3-large, and GPT-4 Turbo (v1106). All model performances are evaluated on the same 10\% held-out test set. See Appendix Table~\ref{tab:classifier-full-results} for the performance results of all 62 trained models. Note that Acc.: Accuracy, F1-M: Macro F1-score, F1-W: Weighted F1-score.}
\label{tab:classifier-results}
\end{table}

\subsubsection{Training and Evaluating Classifiers} To find the best-performing classifier, we systematically trained and evaluated various machine learning models, deep learning models, and LLMs on the three-class classification task, using combinations of text-based video metadata features such as titles, descriptions, transcripts, tags, and comments. Models included Support Vector Machine \cite{cortes1995support}, XGBoost \cite{10.1145/2939672.2939785}, DeBerta-v3-base and DeBerta-v3-large \cite{he2021debertav3}, and GPT-4 Turbo (v1106) \cite{openai2024gpt4technicalreport}. We held out 10\% of the ground-truth dataset as our test set and used accuracy, weighted F1-score, and macro F1-score to evaluate the model performances.
The remaining 90\% of the dataset was employed for training and validation. See Appendix \ref{sec:classifier-training-details} for the input feature descriptions and training procedure. 

We evaluated and trained 62 models with different feature combinations and settings. For brevity, Table \ref{tab:classifier-results} presents the highest performance results achieved by each model. The DeBerta-v3-large model outperformed other models across all performance metrics on the held-out test set, obtaining a test accuracy, weighted F1, and macro F1 of 0.85. See Appendix Table  \ref{tab:classifier-full-results} for the performance results of all 62 models. In Appendix \ref{sec:classifier-performance}, we provide an in-depth discussion on model results and important input features for detecting COVID-19 misinformation in YouTube videos. 

\subsubsection{Remarks.} 
Our best-performing model on the three-class classification task achieves a comparable or even exceeds the performance of other models from prior studies \cite{info:doi/10.2196/49061, 10.1145/3340555.3353763, medina-serrano-etal-2020-nlp, liaw2023younicon}. For example, \citet{papadamou} developed a binary classifier that detected pseudoscientific videos with an accuracy of 0.79 and an F1-score of 0.74. Nevertheless, we acknowledge that this is not a perfect performance, reflecting the complex nature of identifying COVID-19 misinformation within videos and signaling the need for further research efforts. With a 15\% error rate in our classifier, we took additional steps to validate the reliability of our findings due to potential labeling errors (see subsection ``Validation of Results'' in Appendix \ref{sec:validation-analysis}).

\subsection{Handling Videos in Non-English Languages}\label{non-english}
Our classifier, trained exclusively on English videos, was applied only to annotate English videos within the remaining unlabeled portion of the dataset. Non-English videos were manually annotated separately. To identify non-English videos in the remaining dataset, we employed two tools: Google Translate's \textit{Language Detection API} \cite{google_translate} and \textit{langdetect}\footnote{https://pypi.org/project/langdetect/} library to predict the language of the video based on text-based metadata such as the title. For videos flagged as non-English by any of the tools, we manually verified to confirm the video's language. After identifying non-English videos in the remaining dataset, we merged them with the 317 non-English videos from the ground-truth dataset, totaling 784 confirmed non-English videos in the entire dataset. We manually annotated each video using Google Translate and referenced external researchers at our institution fluent in the respective languages.

\section{Quantifying Misinformation Bias}\label{sec:serp-ms}

To quantify the misinformation present in YouTube SERPs, we adopted the misinformation bias score metric from \citet{10.1145/3392854}. The score determines the misinformation bias in a ranked list, giving \textit{more weight }to the annotation labels of \textit{higher-ranked videos} in the calculation: $\frac{\sum^{n}_{r=1}(x_r * (n-r+1))}{\frac{n*(n+1)}{2}}$, where $x$ represents the annotation label of the video, $r$ represents the rank of the video in the SERP (e.g., $r=1$ indicates the top-most video in the SERP), and $n$ represents the total number of videos in the SERP. To conform to the video annotation scale in \citet{10.1145/3392854}, we mapped our 3-class labels from \S \ref{sec:classifier} to a normalized scale of -1, 0, and 1 based on their stance towards COVID-19 misinformation. Videos that oppose COVID-19 misinformation were assigned scores of -1, while those supporting it received a score of 1. However, videos that fell into the merged category, including irrelevant, neutral, and COVID-19 origins labels, do not support nor oppose COVID-19 misinformation. Thus, they\footnote{We also assigned a 0 score to removed videos whose URL was inaccessible ($<$1\% of data), providing a conservative estimate of the misinformation bias in SERPs.} received a score of 0. 
Therefore, the misinformation bias score of a SERP is a continuous value ranging from -1 (all videos oppose misinformation) to +1 (all videos support misinformation). Positive scores indicate a lean toward misinformation, negative scores indicate a lean toward opposing misinformation, and 0 suggests a set of content that neither supports nor opposes misinformation. A higher score suggests a higher prevalence of misinformation in the SERP.

\begin{figure}[!t]
  \centering
\includegraphics[width=0.58\linewidth]{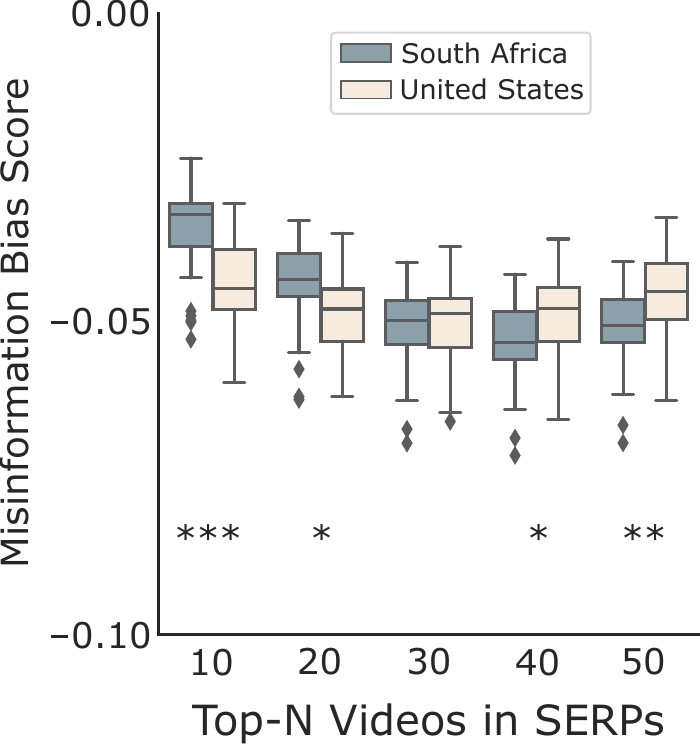}
  \caption{Distribution of the mean misinformation bias scores for the top-10 to top-50 videos in SERPs across the US and SA geolocations. These scores were computed considering the top number of videos (N) in the SERPs. Scores greater than 0 indicate that the SERPs lean toward supporting misinformation, while scores below 0 suggest SERPs lean toward opposing misinformation. Higher scores reflect a greater prevalence of misinformative videos. To compare scores between the US and SA at each level, we performed Mann-Whitney U Tests with test statistics in Appendix Table \ref{tab:top10-50-mann-whitney}. Note that: *\textit{p}$<$0.05; **\textit{p}$<$0.01; ***\textit{p}$<$0.001.}
  \label{fig:top10-50-distribution}
\end{figure}

\section{Results}\label{sec:results}
Here, we compare the prevalence of misinformation in YouTube SERPs across geolocations, topics, and filters. A test of normality revealed that our data is not normal. Thus, we used the non-parametric Mann-Whitney U Test for pairwise comparisons and Kruskal-Wallis tests, followed by the post-hoc Conover-Iman tests with Bonferroni adjustment for multiple comparisons. We provide  details of all significance tests in Appendix \ref{sec:reproducible}. Additionally, we include additional analyses on the misinformation bias scores in search queries and temporal trends of misinformation bias between the US and SA in Appendix \ref{sec:additional-analysis}. Lastly, we conducted validation checks to reinforce the robustness of our findings, accounting for the classifier's 15\% error rate and alternative treatments of the COVID-19 origins class in Appendix \ref{sec:validation-analysis}.

\subsection{Misinformation Bias Across Geolocations}

Figure \ref{fig:top10-50-distribution} displays the distribution of the mean misinformation bias scores for the top-10 to top-50 videos in SERPs between geolocations in the US and SA. 
To get the mean misinformation bias scores, we computed the bias scores considering the top-N videos in the SERPs, averaging across all queries, filters, and bots at a particular geolocation. 

\noindent\textbf{Among the top-N videos, the top-10 videos in the SERPs have the highest misinformation bias scores.} We observed that the top-10 videos in the SERPs have the highest misinformation bias scores for both the US and SA (see Figure \ref{fig:top10-50-distribution}). This suggests that misinformative content is more prevalent in top-10 search results than lower-ranked results. Overall, the misinformation bias scores across the top-N videos in the SERPs are consistently near -0.05, indicating an general trend towards neutral-to-opposing misinformation in both the US and SA. However, we observe that nearly a third of the top-10 search results were misinformative: 31.55\% of the search results supported COVID-19 misinformation, 36.03\% were opposing, and 32.42\% belonged to the remaining classes. 

\noindent\textbf{Bots in SA encountered significantly more misinformative SERPs than bots in the US for the top-10 and top-20 videos.} For the top-10 videos, the effect size \textit{r} was 0.49 (see Appendix Table \ref{tab:top10-50-mann-whitney}), indicating a ``medium effect'' \cite{cohen2013statistical}. This effect size indicates that our observed differences moderately carry practical significance, suggesting that geolocation influences the algorithmic behaviors of YouTube search --- in our case, resulting in statistically significantly more misinformative SERPs in SA compared to the US. We discuss the potential implications of the observed differences in \S \ref{sec:discussion}. 

\noindent\textbf{When considering top-40 and top-50 search results, bots in the US encountered significantly more misinformative SERPs than bots in SA.} This suggests that misinformative videos are still present in the US, but are downranked to lower positions in SERPs, indicating that YouTube's content moderation strategy in the US mitigates misinformation through algorithmic adjustments---downranking such content and reducing their visibility rather than outright removal. However, it is important to note that 95\% of user traffic is directed towards the first page of search results  \cite{web_traffic}, where the top-10 results reside, making it more likely for users to interact with these results than with lower-ranked results. As a result, the difference in the top-40 and top-50 results may have less impact on user exposure compared to the top-10 results, where SA users are more likely to encounter misinformation. Since video viewers are most likely to engage with the top-10 videos in SERPs, our subsequent analysis focuses on these search results.

\begin{figure}[t!]
  \centering
  \includegraphics[width=0.95\linewidth]{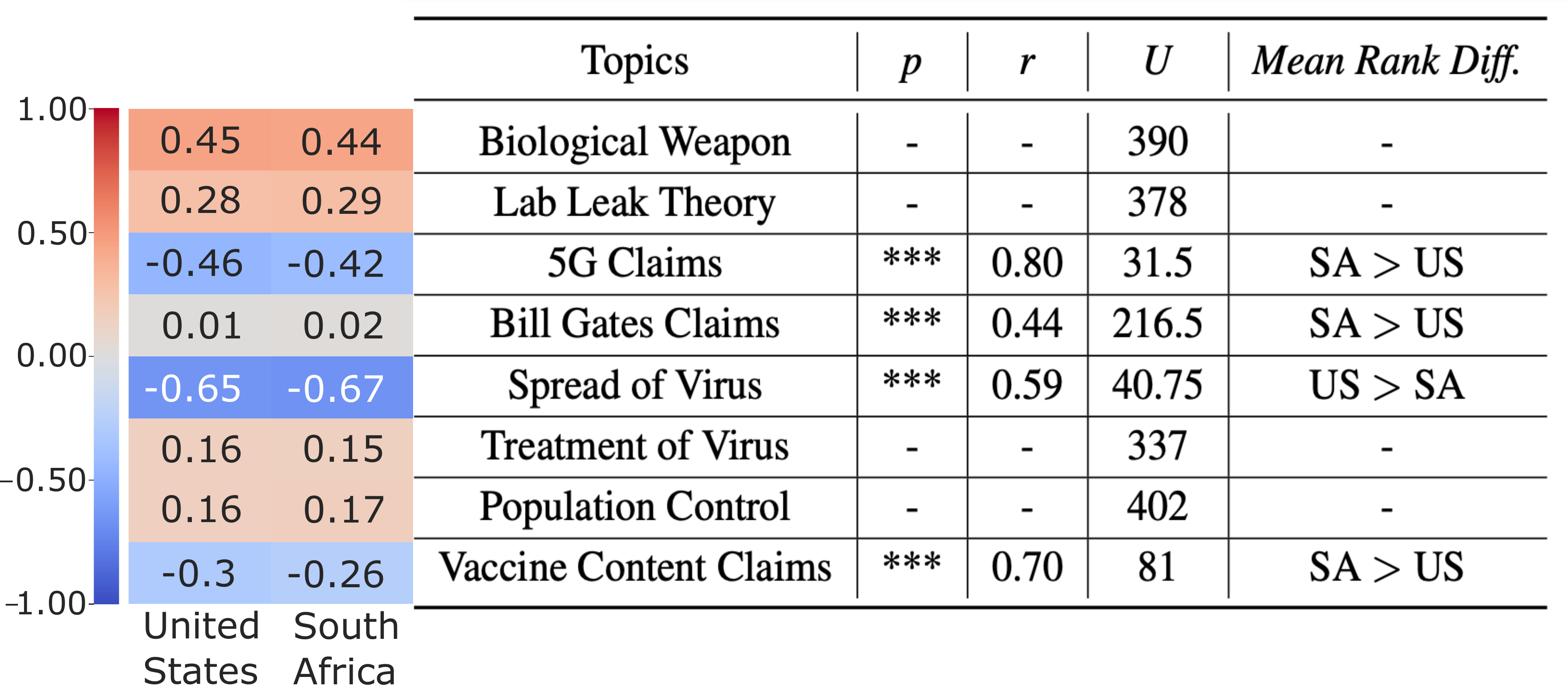}
  \caption{For each topic, we indicate the average misinformation bias scores of the top-10 search results between the US and SA (heatmap) and conduct a Mann-Whitney U Test to compare these bias scores between the two countries. We denote the p-value (\textit{p}), Mann-Whitney effect size (\textit{r}), \textit{U}-value, and the mean rank difference. For example, for the ``5G Claims'' topic, SA $>$ US in the ``Mean Rank Diff.'' column indicates that bots in SA received more misinformative videos in the top-10 search results than bots in the US. 
  Note that: *\textit{p}$<$0.05; **\textit{p}$<$0.01; ***\textit{p}$<$0.001.}
  \label{fig:topic_diff}
\end{figure}

\subsection{Misinformation Bias Within Each Country}
How do the geolocations within each country influence the prevalence of misinformation in YouTube SERPs?
To answer, we conducted a Kruskal-Wallis H Test to examine differences in misinformation bias scores across geolocations within each country. The test revealed a significant difference for US geolocations (KW H(2, N$=$30)=6.98, \textit{p}$<$0.01, $\eta^2$=0.18), but no significant difference within SA. The large effect size for the US ($\eta^2$=0.18) \cite{kruskal-effect} suggests that the observed differences carry practical significance, indicating that geolocation may influence the prevalence of misinformation in YouTube SERPs in the US. We conducted a post-hoc Conover-Iman test with Bonferroni adjustment, revealing that bots in Houston received a higher prevalence of misinformation in their search results than in Los Angeles. We discuss the implications in \S \ref{sec:discussion}.

\subsection{Misinformation Bias in Topics}
\begin{figure}[t!]
  \centering
  \includegraphics[width=0.98\linewidth]{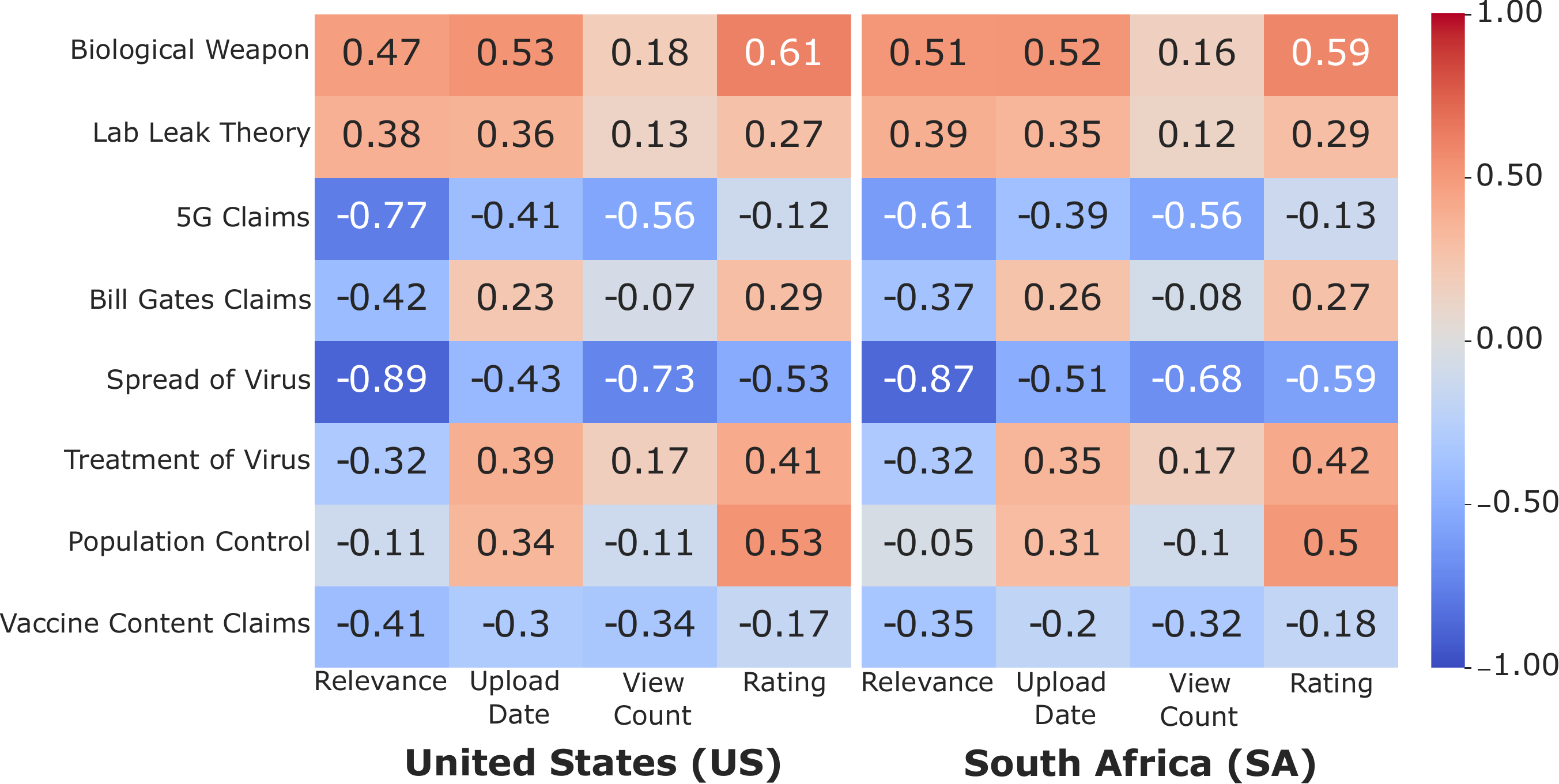}
  \caption{Mean misinformation bias scores for the top-10 search results across all 8 topics and 4 search filters between the US and SA. Note that ``Relevance'' is YouTube's default sorting filter for search results.}
  \label{fig:heatmap_us_sa}
\end{figure}

Figure \ref{fig:topic_diff} shows the mean misinformation bias scores and the Mann-Whitney U Test results between the US and SA across 8 topics. The scores in the heatmap were computed by averaging across each topic's constituting queries, search filters, bots within a country, and the 10-day experimental period. Figure \ref{fig:heatmap_us_sa} shows the misinformation bias scores across the topics, search filters, and countries.

\noindent\textbf{Bots in SA encountered significantly more misinformative SERPs than bots in the US for 3 topics.} In Figure \ref{fig:topic_diff}, the topics ``5G claims'' (\textit{r}=0.80), ``Bill Gates Claims'' (\textit{r}=0.44),  and ``Vaccine Content Claims'' (\textit{r}=0.59) exhibited statistically significant differences in misinformation bias between the US and SA, indicating medium to large effect sizes. These effect sizes imply practical significance, suggesting that users in SA may encounter significantly more misinformative SERPs than users in the US for these topics. Of particular concern is the presence of these differences in topics related to public health, such as ``Vaccine Content Claims.`` In contrast, bots in the US encountered significantly more misinformative SERPs than bots in the SA for a single topic---``Spread of Virus.'' We discuss the implications of these findings in \S\ref{sec:discussion}.

\noindent\textbf{Topics ``Biological Weapon'' and ``Lab Leak Theory'' exhibited positive scores across all search filters in both countries} (see Figure \ref{fig:heatmap_us_sa}).  Given that positive scores indicate that the SERPs lean toward misinformative content, this finding suggests the problematic nature of these topics, with bots exposed to misinformative SERPs regardless of their geolocations or the search filters used.

\noindent\textbf{Topics ``5G Claims,'' ``Spread of Virus,'' and ``Vaccine Content Claims'' displayed negative scores across all search filters in both countries.} This suggests that SERPs for these topics generally oppose misinformation in both the US and SA. This result may be attributed to YouTube's targeted content moderation efforts on these topics \cite{youtube_statement}, potentially highlighting some successes in combating misinformation on its platform.

\noindent\textbf{Three topics showed negative scores in SERPs sorted by ``Relevance'' and ``View Count,'' but positive scores when sorted by ``Upload Date'' and ``Rating.''} We observe this pattern for ``Bill Gates Claims, ``Treatment of Virus,'' and ``Population Control'' topics, indicating that bots encounter more misinformative SERPs when searching for newly uploaded or highly rated videos for such topics.

\subsection{Misinformation Bias in Search Filters}
\begin{figure}[t!]
  \centering
  \includegraphics[width=0.95\linewidth]{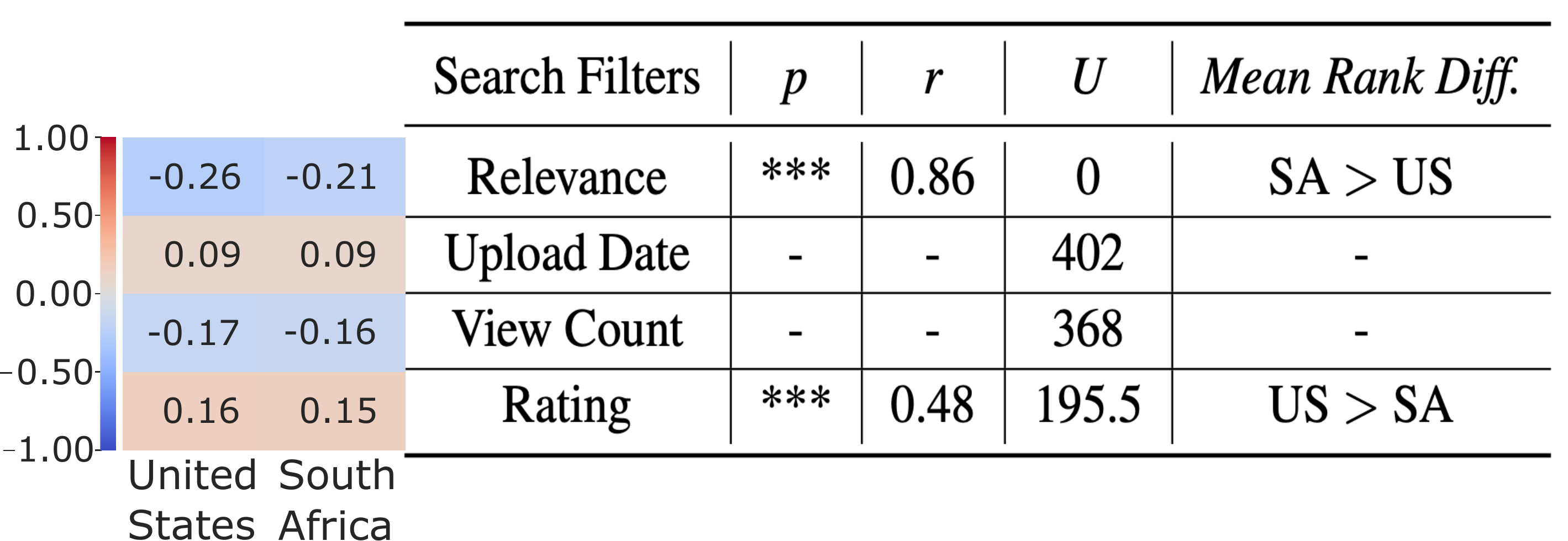}
  \caption{For each search filter, we indicate the average misinformation bias scores of the top-10 search results between the US and SA (heatmap) and conduct a Mann-Whitney U Test to compare these bias scores between the two countries.
  *\textit{p}$<$0.05; **\textit{p}$<$0.01; ***\textit{p}$<$0.001.}
  \label{fig:filter_diff}
\end{figure}

Figure \ref{fig:filter_diff} depicts the mean misinformation bias scores and the Mann-Whitney U Test results between the US and SA across 4 search filters. 

\noindent\textbf{Both ``Upload Date'' and ``Rating'' filters showed positive scores in both countries.} This indicates that SERPs sorted for highly rated (more likes than dislikes) or newly uploaded videos lean towards misinformative content. Our findings suggest that highly rated and newly uploaded videos tend to feature misinformation. This observation suggests that YouTube may not prioritize its content moderation efforts toward recently uploaded or highly rated videos.

\noindent\textbf{In contrast, ``Relevance'' and ``View Count'' filters showed negative scores in both countries.} This suggests that SERPs sorted by relevant or most-viewed videos lean towards opposing misinformation. This may suggest that YouTube's content moderation efforts prioritize relevant and highly viewed videos, which are likely to be surfaced by the search engine or have gained viewer attention. 

\noindent\textbf{Bots in SA received significantly more misinformative SERPs than bots in the US when sorting the results by ``Relevance.''} As shown in Figure \ref{fig:filter_diff}, the effect size of \textit{r=0.86} indicates a large effect, suggesting that users in SA may encounter more misinformative SERPs than their US counterparts when using the default ``Relevance'' filter.

\noindent\textbf{However, bots in the US received significantly more misinformative SERPs than those in SA when sorting the results by ``Rating.''} The effect size \textit{r=0.48} indicates a medium effect. While this effect size is not as pronounced as when sorting by ``Relevance'' (\textit{r=0.86}), it suggests that the observed differences have moderate practical significance. It is important to note that YouTube's default filter is ``Relevance,'' suggesting that users are more likely to engage with SERPs sorted by this criterion. Consequently, the observed differences with the ``Rating'' filter may have a lesser impact on users compared to the ``Relevance'' filter, where bots in SA received significantly more misinformative SERPs than bots in the US. 

\section{Discussion}\label{sec:discussion}

\subsubsection{COVID-19 Misinformation on YouTube:}  Auditing YouTube's search engine for COVID-19 misinformation is urgently needed to ensure algorithmic accountability, protect public health, and develop more responsible web. Despite being designated a global pandemic in 2020, COVID-19 remains a public health threat \cite{still_pandemic}. While YouTube has been a valuable tool for finding health information during the pandemic, it has also disseminated harmful misinformation, impeding public health efforts and vaccine hesitancy \cite{who_infodemic}.

Our audit, conducted in early 2023, uncovered that 31.55\% of the top-10 search results contained COVID-19 misinformation. This percentage aligns with a 2020 study \cite{Lie002604}, which identified that over 25\% of the most-viewed YouTube videos contain misleading information. Despite the pandemic being declared in 2020, our findings indicate that COVID-19 misinformation remains pervasive in YouTube search results three years later.

Notably, topics like ``Biological Weapon'' and ``Lab Leak Theory'' were consistently contaminated with COVID-19 misinformation across all search filters in both countries (Figure \ref{fig:heatmap_us_sa}). These results indicate the problematic and global nature of these misinformation topics, highlighting the urgent need for YouTube to enhance its content moderation in these areas. On the other hand, topics ``5G Claims,'' ``Spread of Virus,'' and ``Vaccine Content Claims'' contained opposing search results across all filters in both countries. This finding may be attributed to YouTube's content moderation policies and its commitment to reducing misinformative content relating to 5G conspiracy, vaccines, and information that contradicts medical authorities \cite{youtube_statement, youtube_statement2, google_misinfo}. Overall, our study suggests that while YouTube may be successful in moderating content on some topics, it still has considerable work to do in moderating COVID-19 misinformation in its search engine.  

\subsubsection{Misinformation in Emerging and Highly-Rated Content:} We observed an alarming trend in both countries for certain filters such as ``Upload Date'' and ``Rating.'' As illustrated in Figure \ref{fig:filter_diff}, the misinformation bias scores were positive for these filters, indicating that emerging and highly rated videos tend to feature misinformation. This is problematic, as users may place trust in videos highly rated by others, and users seeking the latest information about the pandemic may be at heightened risk of encountering misinformation.
On August 25th, 2023, the European Union's Digital Services Act went into effect, placing responsibility on online search engines, including YouTube, for the misinformative content on the platforms \cite{eu_dsa2}. In light of these regulations, our study provides valuable insights for YouTube to enhance its content moderation practices, especially in addressing emerging or highly rated content that may contribute to the spread of COVID-19 misinformation and, consequently, impact public health. 

\subsubsection{Contrasting Algorithmic Behaviors of YouTube Search in Different Geolocations:} 
We observe several instances of statistically significant disparities, with bots in SA encountering significantly more misinformative SERPs than bots in the US on YouTube, particularly within the top-10 search results (p$<$0.001, \textit{r}=0.49) and search results sorted by the ``Relevance'' filter (p$<$0.001, \textit{r}=0.86). Since users are likely to engage with these search results (see \S\ref{sec:results}), this suggests that users in SA may encounter significantly more misinformative SERPs than users in the US. Considering the established importance of YouTube for finding health-related information \cite{KHATRI2020101636}, exposing users in SA to more misinformation in their daily YouTube usage may increase their personal health risks by potentially negatively influencing their beliefs, health practices, and decisions. 

Furthermore, our results indicate that bots in SA were exposed to significantly more misinformative SERPs compared to bots in the US for the topics ``5G Claims,'' ``Bill Gates Claims,'' and ``Vaccine Content Claims'' (Figure \ref{fig:topic_diff}). 
Notably, South Africa experienced distressing incidents during the pandemic, such as the burning of 5G towers fueled by a conspiracy theory linking them to the spread of COVID-19 \cite{5g_towers_sa}. Concurrently, South Africa grappled with many public health challenges, including a vaccine supply shortage and a deep distrust of medical authorities contributing to vaccine hesitancy \cite{vaccine_skepticism}. Given these circumstances, the heightened exposure of SA users to misinformation compared to US users on these topics could potentially exacerbate societal issues in SA, including vaccine hesitancy and undermining public health efforts. 

Within the US, our analysis revealed that bots in Houston, Texas, received significantly more misinformative SERPs than those in Los Angeles, California. Prior work has shown that conservatives were more susceptible to believing falsehoods and conspiracy beliefs \cite{conservative}. Since Houston is in a conservative-leaning state (Texas), promoting significantly more misinformative search results to conservative-leaning users could exacerbate harm by reinforcing or fostering beliefs in COVID-19 misinformation.

However, political leaning alone is unlikely to explain the differences observed within the US. Other factors, such as regional search behaviors, public health policies, and online media use policies may also play a role. The observed geographic differences within the US, as well as between the US and SA, highlight important societal and policy implications for platforms like YouTube. These platforms should assess whether current moderation policies sufficiently address geographic disparities in the prevalence of COVID-19 misinformation in search results. To mitigate these disparities, platforms could analyze region-specific search behaviors and trends to adjust rankings or visibility for content likely to contain misinformation. Additionally, regionally tailored counter-misinformation efforts—such as promoting public health campaigns in high-risk areas—could help curb misinformation across diverse regions.

\subsubsection{The Challenges and Importance of Conducting Audits in the Global South:} While algorithmic audits have been predominantly focused on the Global North contexts, there has been limited exploration into audits concerning the Global South. Our research explores the less-researched Global South context by conducting the first large-scale geolocation-based audit of YouTube search for COVID-19 misinformation between the US and SA. 

However, we must acknowledge the challenge posed by the limited availability of affordable proxies, VPNs, and virtual machines in Global South countries. This makes large-scale audit experiments in the Global South either cost-prohibitive or, in some cases, logistically impossible. Addressing this issue underscores the pressing need for enhanced technological access and affordability to facilitate more \textit{extensive} research initiatives within the Global South, such as conducting audits in multiple countries in the region.

Disparate algorithmic behaviors based on geolocation raise concerns regarding fairness, especially when returning significantly more misinformative SERPs in one region over another. 
Our findings serve as an important call to action for YouTube to regulate its algorithmic behavior consistently to ensure equitable and accurate information dissemination across different regions.
This proactive approach is crucial to facilitate the development of a more responsible and fair video search engine that preserves factual and accurate information irrespective of the user's geolocation. 

Historically, past audits have successfully generated awareness about these issues, creating pressure on platforms to improve their algorithm \cite{mit_actionable}. Similarly, we hope our work drives meaningful changes on YouTube and inspires future audits in the Global South.

\section{Limitations and Future Directions}
Our work is not without limitations. Our audit used only English search queries to maintain consistency in search query language across both countries. Future studies can audit YouTube's search results by utilizing queries in diverse regional and local languages within the countries. In addition, our study focuses on YouTube's search engine; future studies can investigate the impacts of geolocation on the recommendation systems on the home page and video page. As the COVID-19 pandemic remains a public health threat globally, our study focuses on COVID-19 misinformation on YouTube. However, other domains, such as climate change misinformation, warrants attention---our methods, including the experimental setup, can be adapted to investigate misinformation beyond COVID-19.

In our study, we employ sock-puppet bots to emulate the actions of real users, which avoids contaminating real users' search and recommendation results (explained further in \S \ref{sec:ethics}). However, it is important to note that while we can analyze what happens when a hypothetical user searches on YouTube, we cannot determine the prevalence of misinformation for \textit{real} users in South Africa compared to the United States. To address this limitation, future work could ethically and carefully conduct \textit{crowdsourced} audits \cite{10.1145/3544548.3580846} to gather real user data and compare misinformation prevalence in these different geographic regions.

Given the challenges of conducting audits in the Global South, our audit focuses on a country from the Global North and one from the Global South. Future works should conduct audits in multiple countries within the Global South. Our work uses the misinformation bias score \cite{10.1145/3392854}, capturing the amount of misinformation and the video's rank but not its relevance. Future works can use metrics like the Convex Aggregated Measure \cite{10.1145/3121050.3121072}, which uses relevance and credibility in ranked lists.

Although we showed in \S \ref{sec:classifer_training} that our classifier performs comparably to YouTube models tackling related issues, we acknowledge that our classifier has an error rate of 15\%, potentially influencing downstream misinformation analysis. We partly address this limitation by conducting additional analyses, which provided further evidence of the reliability of our findings despite the error rate of our classifier (see subsection ``Validation of Results'' in Appendix \ref{sec:validation-analysis}).

\section{Conclusion}
In this study, we conducted an audit on YouTube comparing the prevalence of COVID-19 misinformation in SERPs between the US and SA. Our findings revealed that overall, 31.55\% of the top-10 search results contained COVID-19 misinformation. Among these top-10 search results, we found that bots in SA encountered significantly more misinformative SERPs than bots in the US. Overall, our study highlights the potential need to regulate algorithmic behavior consistently across different regions and calls for future research efforts to consider contexts in the Global South. 

\section*{Acknowledgements}
Mitra and Jung were partially supported by National Institute of Health's NIDA grant DA056725-01A1. Jung was also supported by the Mary Gates Endowment and the National Science Foundation grant \#2128642. We thank the Poynter Fact-Checking Innovation Initiative award for supporting this work, PesaCheck for the fact-checked video dataset, and AMT workers for their annotations.

\begin{small}
    \bibliography{aaai25}
\end{small}

\section*{Ethical Statement}\label{sec:ethics}
We took several steps to minimize potential harm from our audit experiments. First, we refrained from recruiting real-world users in the data collection phase of our audit experiments. According to \citet{10.1145/3392854}, searching misinformative queries could contaminate users' searches and recommendations, which may have long-term consequences regarding what videos are recommended to participants by YouTube's algorithm. Thus, in our audit experiments, we employed sock-puppet bots that emulated the actions of actual users. Second, to set up our bots' IP address geolocations, we used proxies from \textit{IPRoyal}, which ethically source their proxies from consenting individuals who willingly share their bandwidth in exchange for compensation. Furthermore, we did not use the IP addresses to identify the individuals sharing their bandwidth and deleted all IP address information after the experiment. 

Additionally, we acknowledge the potential harm of exposing AMT workers to misinformative content on YouTube. We took four measures to mitigate such risks and impacts to annotators. First, we explicitly detailed the labeling task instructions at the beginning of the Qualification Test, providing an early warning to the annotators regarding possible misinformative content in the video they are about to watch. We also added the same instructions and warning in the actual task itself. Second, to clarify and help annotators understand what misinformation is in their annotation task, we explicitly provided several examples of videos containing COVID-19 misinformation and also included a reference to Google's COVID-19 medical misinformation policy \cite{google_misinfo}, which lists out debunked misinformation claims surrounding the COVID-19 pandemic. Third, to address any potential distress, concerns, or confusion, we included our lab email in the annotation instructions to provide a clear and easy way for annotators to contact us. Fourth, annotators could quit the task anytime and get paid for their work.

\section*{Paper Checklist}

\begin{enumerate}

\item For most authors...
\begin{enumerate}
    \item  Would answering this research question advance science without violating social contracts, such as violating privacy norms, perpetuating unfair profiling, exacerbating the socio-economic divide, or implying disrespect to societies or cultures?
    \answerYes{Yes. Our work fills a critical gap in algorithmic audit research, providing the first large-scale geolocation-based comparative audit of YouTube search for COVID-19 misinformation between a country in the Global North and one in the South. Our work highlights fairness concerns related to the differing algorithmic behaviors based on geolocation and serves as an important call to action for YouTube to regulate its algorithmic behaviors consistently across all regions.}
  \item Do your main claims in the abstract and introduction accurately reflect the paper's contributions and scope?
    \answerYes{Yes. We carefully reviewed the claims in the abstract and Introduction to ensure that they accurately reflect our contributions and project scope.}
   \item Do you clarify how the proposed methodological approach is appropriate for the claims made? 
    \answerYes{Yes. Please refer to Sections 3-5. We justify the appropriateness of our methods, such as our search query curation and experimental design, by referencing prior works with similar tasks/objectives. For our analysis results, we justify the use of the misinformation bias score metric based on its usage in previous studies quantifying misinformation in search engine result pages. Our classifier's performance is quantitatively validated by comparing the classifier labels with human annotations, and we contextualize this performance by referencing prior works (see subsection ``Remarks'' under Section 5). Overall, we justify our methodological approach throughout the paper.}
   \item Do you clarify what are possible artifacts in the data used, given population-specific distributions?
    \answerYes{Yes. Please refer to Section 3 and Appendix Tables 3-4, where we describe the characteristics and sample data points from the \textit{PesaCheck} dataset, which we used to derive the set of search queries for the audit experiment.}
  \item Did you describe the limitations of your work?
    \answerYes{Yes. We discuss the limitations of our work in Section 9 ``Limitations and Future Directions.''}
  \item Did you discuss any potential negative societal impacts of your work?
    \answerYes{Yes. Refer to Section 11 ``Ethics Statement'' where we discuss the potential negative impacts of our work and the steps we took to mitigate them.}
  \item Did you discuss any potential misuse of your work?
    \answerNA{NA}
  \item Did you describe steps taken to prevent or mitigate potential negative outcomes of the research, such as data and model documentation, data anonymization, responsible release, access control, and the reproducibility of findings?
    \answerYes{Yes. In Section 11 ``Ethics Statement,'' we followed best practices to mitigate potential harm to human annotators. We ethically sourced proxies from consenting individuals and properly deleted all IP addresses after the experiment. To ensure responsible release, we will release the classifier on HuggingFace and data on a Github repository (see footnote 1). For reproducibility, we thoroughly document the significance testing details in Appendix Section I.}
  \item Have you read the ethics review guidelines and ensured that your paper conforms to them?
    \answerYes{Yes, I read the ethics review guidelines and ensured that our paper conforms to them to the best of our capabilities.}
\end{enumerate}

\item Additionally, if your study involves hypotheses testing...
\begin{enumerate}
  \item Did you clearly state the assumptions underlying all theoretical results?
    \answerNA{NA}
  \item Have you provided justifications for all theoretical results?
    \answerNA{NA}
  \item Did you discuss competing hypotheses or theories that might challenge or complement your theoretical results?
   \answerNA{NA}
  \item Have you considered alternative mechanisms or explanations that might account for the same outcomes observed in your study?
    \answerNA{NA}
  \item Did you address potential biases or limitations in your theoretical framework?
    \answerNA{NA}
  \item Have you related your theoretical results to the existing literature in social science?
    \answerNA{NA}
  \item Did you discuss the implications of your theoretical results for policy, practice, or further research in the social science domain?
    \answerNA{NA}
\end{enumerate}

\item Additionally, if you are including theoretical proofs...
\begin{enumerate}
  \item Did you state the full set of assumptions of all theoretical results?
    \answerNA{NA}
	\item Did you include complete proofs of all theoretical results?
    \answerNA{NA}
\end{enumerate}

\item Additionally, if you ran machine learning experiments...
\begin{enumerate}
  \item Did you include the code, data, and instructions needed to reproduce the main experimental results (either in the supplemental material or as a URL)?
    \answerYes{Yes, in the ``Data Set'' entry accompanying the submission, we provide in-depth instruction and data needed to reproduce our main experimental results. We will also make this public via Github Repository.}
  \item Did you specify all the training details (e.g., data splits, hyperparameters, how they were chosen)?
    \answerYes{Yes. In the Appendix Section G  ``Classifier Training Details,'' we discuss the training details, such as feature descriptions, training procedures, prompt design considerations, etc.}
     \item Did you report error bars (e.g., with respect to the random seed after running experiments multiple times)?
    \answerNA{NA}
	\item Did you include the total amount of compute and the type of resources used (e.g., type of GPUs, internal cluster, or cloud provider)?
    \answerYes{Yes. In the Appendix Section G ``Classifier Training Details,'' we discuss the type of GPU employed to train the deep learning models. Most of our training did not require intensive computing or an external cloud provider.}
     \item Do you justify how the proposed evaluation is sufficient and appropriate to the claims made? 
    \answerYes{Yes. Please refer to the subsection ``Training and Evaluating Classifiers'' (under Section 5 ``Labeling YouTube Videos''), where we quantitatively validate our classifier by comparing machine-generated and human-generated labels. We contextualize our model performance with the existing COVID-19 misinformation detection literature. Further, in Appendix Section \ref{sec:validation-analysis} subsection ``Validation of Results,'' we validate the reliability of our findings due to potential labeling errors by the classifier.}
     \item Do you discuss what is ``the cost`` of misclassification and fault (in)tolerance?
    \answerYes{Yes. In the Section ``Limitations and Future Directions,'' we acknowledge the error rate of our classifier, which may potentially influence our downstream analysis. To mitigate such impacts, we conducted further validation in the Appendix Section \ref{sec:validation-analysis} subsection ``Validation of Results,'' which provides further evidence of the reliability of our findings despite the misclassification error rate.}
  
\end{enumerate}

\item Additionally, if you are using existing assets (e.g., code, data, models) or curating/releasing new assets, \textbf{without compromising anonymity}...
\begin{enumerate}
  \item If your work uses existing assets, did you cite the creators?
    \answerYes{Yes. In the subsection ``Training and Evaluating Classifiers,'' we cite the creators of all the machine learning models employed in our study. Additionally, we cite the creators of the COVID-19 QA datasets and \textit{KeyBERT} in the Section ``Curating Topics and Search Queries.''}
  \item Did you mention the license of the assets?
    \answerNo{No, we did not due to space constraints. The \textit{all-mpnet-base-v2} model,\footnote{https://huggingface.co/sentence-transformers/all-mpnet-base-v2} Selenium Python Library,\footnote{https://www.selenium.dev/documentation/about/copyright/} and COVID-19 QA dataset from \citet{oller2020covidqa} are under the apache-2.0 license, and the remaining models, libraries, and tools employed in the study are released under the MIT license.}
  \item Did you include any new assets in the supplemental material or as a URL?
    \answerYes{Yes, we included our dataset in the ``Data Set'' entry accompanying the submission.}
  \item Did you discuss whether and how consent was obtained from people whose data you're using/curating?
    \answerNo{Since we collected publicly available data on YouTube, explicit consent from people whose data we are curating is not required. Analyzing this retrospective data did not constitute human subjects research, and thus, informed consent is not required from the video creators.}
  \item Did you discuss whether the data you are using/curating contains personally identifiable information or offensive content?
    \answerYes{Yes, given our data's offensive and misleading nature, we explicitly provide a content warning after the abstract.}
\item If you are curating or releasing new datasets, did you discuss how you intend to make your datasets FAIR?
\answerNo{No, we did not due to space constraints. We will publish the provided dataset onto a publicly-facing Github repository (F, A) in .csv format (I). The dataset will be released under the MIT license (R).}
\item If you are curating or releasing new datasets, did you create a Datasheet for the Dataset? 
\answerYes{Yes, in the ``Data Set'' entry accompanying the submission, we provide a detailed README.MD file that provides an overview and data description of the datasets. We will also release the same README.MD file in the public Github repository.}
\end{enumerate}

\item Additionally, if you used crowdsourcing or conducted research with human subjects, \textbf{without compromising anonymity}...
\begin{enumerate}
  \item Did you include the full text of instructions given to participants and screenshots?
    \answerYes{Yes, Figures 9-13 contain the screenshots and instruction texts given to participants.}
  \item Did you describe any potential participant risks, with mentions of Institutional Review Board (IRB) approvals?
    \answerYes{Yes, we described the potential participant risks in the Section ``Ethics Statement.'' Following the best practices from a prior audit work \cite{papadamou}, we took extensive steps to mitigate participant risks and ensure informed consent when designing our task.}
  \item Did you include the estimated hourly wage paid to participants and the total amount spent on participant compensation?
    \answerYes{Yes. We included the estimated hourly wage paid to participants and the amount spent per participant annotation in Appendix Section F.}
   \item Did you discuss how data is stored, shared, and deidentified?
   \answerYes{Yes. We describe how we processed the annotations in the Appendix Subsection ``YouTube Annotation Task.'' We employed a majority agreement to arrive at a final label for a YouTube video. Since only the final labels are included in our dataset, no identifiable information about the annotators is shared.}
\end{enumerate}

\end{enumerate}

\appendix

\begin{table}[ht]
\small
\centering
\begin{tabular}{ll}
\toprule
\multicolumn{2}{c}{\textbf{Countries}}   \\ \midrule
Australia & Algeria\\
Canada & Bangladesh\\
Hungary & China\\
Ireland & Ethiopia\\
Italy & India\\
Japan & Iraq\\
Republic of Korea & Jamaica\\
New Zealand & Kenya\\
Singapore & Malaysia\\
United Kingdom & Morocco\\
United States & Nigeria\\
Pakistan & Philippines\\
Qatar & South Africa\\
Sri Lanka & Uganda\\
United Arab Emirates\\
\bottomrule
\end{tabular}
\caption{\textit{PesaCheck}'s fact-checked dataset contains 362 videos, which originate from channels based in 29 different countries. The 29 countries span both the Global North and the Global South, with 18 of the 29 countries being associated with the Global South.}
\label{tab:pesacheck-countries}
\end{table}

\begin{table*}[ht]
\small
\centering
\begin{tabular}{lp{10cm}}
\toprule
\textbf{Misinformation Topics} & \textbf{Sample YouTube Videos (Video Title and ID)} \\
\midrule
Biological Weapon & 
  PG-18 Reasons \& evidence that I infer COVID-19 as the biological weapon made by Chinese Government (\textit{Video ID: 2uDpARUBPEE}) \\
\cmidrule{2-2}
Lab Leak Theory & 
  Russian Flu of 1977 Linked to Lab Leak \& Live Vaccine Trials Startling Similarities to COVID-19 (\textit{Video ID: us8gd\_XY2fY}) \\
\cmidrule{2-2}
5G Claims & 
  Why The Coronavirus Test Is Likely The Bill Gates 5G Google Chip (\textit{Video ID: 02RVwXRsAHg}) \\
\cmidrule{2-2}
Bill Gates Claims & 
  Rev Chris Okotie: THE COVID 19 MYSTERY (\textit{Video ID: 8uPbf7T0KbQ}) \\
\cmidrule{2-2}
Spread of Virus & 
  How to Stay Safe and Prevent the Spread of COVID 19 (\textit{Video ID: 8SDneTSzB5E}) \\
\cmidrule{2-2}
Treatment of Virus & 
  COVID-19 SELF TREATMENT: How to determine which remedies to use (\textit{Video ID: 92DWaf6qmWM}) \\
\cmidrule{2-2}
Population Control & 
  W.H.O WHISTLEBLOWER: Depopulation, total control, Perpetual fear and perpetual vaccination. (\textit{Video ID: vD6IcaAFORM}) \\
\cmidrule{2-2}
Vaccine Content Claims & 
  COVID 19 VACCINES ARE A GIFT FROM ABORTED BABIES? CIVIL RIGHTS FOR CATHOLICS? (\textit{Video ID: kRTv8G2VI1M}) \\
\bottomrule
\end{tabular}
\caption{The 8 globally-persistent COVID-19 misinformation topics identified by \textit{PesaCheck}. For each topic, we provide a sample YouTube video from \textit{PesaCheck}'s dataset.}
\label{tab:topics-sample}
\end{table*}

\section{Model Fine-tuning}\label{sec:finetuning}
We selected the Sentence Transformer model \textit{all-mpnet-base-v2} due to their best performance generating sentence embeddings across various tasks \cite{song2020mpnet}. However, \textit{all-mpnet-base-v2} was not trained on COVID-19-related texts. \citet{scialom2022finetuned} demonstrated that language models can be “continual learners” of new domains via self-supervised training. Thus, we continued training the model using a self-supervised contrastive learning framework (i.e. multiple-negative-ranking-loss) on COVID-19 question-answering (QA) datasets \cite{oller2020covidqa, covid-qa, tang2020rapidly}, totaling 3,296 QA pairs, to generate more meaningful sentence embeddings on the COVID-19-related texts. We randomly partitioned the dataset into 80\% training, 10\% validation, and 10\% testing sets. To avoid overfitting, we trained the model until the training loss decreased further relative to the validation loss. Our best model used an Adam optimizer with a learning rate of 2e-5, 3 epochs, 32 batch size, and a warmup step for 10\% of total steps. We found that our fine-tuning worked. The finetuned model had a test loss of 0.7466 and the original pre-trained model had a test loss of 1.7722, indicating that the finetuned model generated more meaningful sentence embeddings on COVID-19 pandemic-related texts. 

\section{Measuring Geolocation-based Personalization}\label{sec:personalization_metric}
To measure geolocation-based personalization, we used the Jaccard index to calculate the similarity of two SERPs. Previous audits used the Jaccard index to measure personalization in web searches \cite{10.1145/2815675.2815714,  10.1145/3411764.3445250}. The Jaccard index determines the similarity between two lists -- 1 indicates that both lists have identical elements, while 0 indicates that both have completely different elements. 

We formalize the metric to compute the geolocation-based personalization between two geolocations, location $x$ and location $y$. We indicate the SERPs collected by a treatment bot at location $x$ as $\text{SERP1}_x$ and its corresponding twin bot as $\text{SERP2}_x$. First, we establish the noise level at each geolocation by calculating the Jaccard index of SERPs collected by the identical twins at the same geolocation (Equation \ref{eq:noise}). Ideally, these twin bots should have a Jaccard index of 1 because these are identical “users” searching for the same query at the same time and geolocation; however, even after controlling for all known sources of noise, there may be some sources of noise that we are not aware of. Second, we take the higher noise level (i.e. lower Jaccard’s index) between the geolocations to establish a baseline noise and obtain a more conservative estimate of geolocation-based personalization between the two geolocations (Equation \ref{eq:baseline}). Third, we compute the difference in the SERPs between the two geolocations by calculating the Jaccard's index of SERPs collected by the treatment bot at each geolocation (Equation \ref{eq:difference}). Finally, we calculate the difference between the baseline noise level and Equation \ref{eq:difference} -- the additional difference beyond the baseline noise can then be attributed to geolocation-based personalization, denoted as \texttt{GBP} in Equation \ref{eq:personalization}. 

The \texttt{GBP} metric ranges from 0 to 1 and indicates how much of the differences between the SERPs from location $x$ and location $y$ can be attributed to geolocation-based personalization. A zero indicates that the difference between the treatments did not exceed the baseline noise, while a positive value indicates that the difference in the SERPs can be attributed to geolocation-based personalization. 

\begin{equation}\label{eq:noise}
    \texttt{Noise}(x) = \texttt{Jaccard}(\text{SERP1}_x, \text{SERP2}_x)
\end{equation}
\begin{equation}\label{eq:baseline}
    \texttt{Baseline}(x, y) = \texttt{min}(\texttt{Noise}(x), \texttt{Noise}(y))
\end{equation}
\begin{equation}\label{eq:difference}
    \texttt{Diff.}(x, y) = \texttt{Jaccard}(\text{SERP1}_x, \text{SERP1}_y)
\end{equation}
\begin{equation}\label{eq:personalization}
    \texttt{GBP}(x, y) = \texttt{Baseline}(x, y) - \texttt{Diff.}(x, y)
\end{equation}

\begin{table*}[ht]
\centering
\begin{tabular}{p{3.5cm}p{13cm}}
\toprule
\textbf{Misinformation Topics} & \textbf{Search Queries} \\
\midrule
Biological Weapon & 
  Biological Weapon, CCP virus, man-made virus, China virus, fact chinesemade virus, revelations gravitas china virus coverup \\
\cmidrule{2-2}
Lab Leak Theory & 
  lab leak theory, WHO coverup, china scientist create virus, North Carolina lab in US, Kungflu, evidence virus lab \\
\cmidrule{2-2}
5G Claims &
  5g bad effect, 5g conspiracy, 5g and covid19 link, the dangers of 5g radiation, why 5g testing cause corona, dangers 5g \\
\cmidrule{2-2}
Bill Gates Claims &
  destroy Africa, depopulate the world, vaccine testing africa, bill gates vaccine chip, bill gates exposed, happened vaccine devil \\
\cmidrule{2-2}
Spread of Virus &
  sanitize, dogs and cats, Covid 19 Spread And Precautions, precaution for pets, spread of covid by a bat, social spread \\
\cmidrule{2-2}
Treatment of Virus &
  sesame oil, garlic, herbs, local concoctions, treatment covid19 government, dealing vulnerable population \\
\cmidrule{2-2}
Population Control &
  population control, mass murder, plandemic, nuremberg code, brainwashing, vaccines depopulation \\
\cmidrule{2-2}
Vaccine Content Claims &
  MRNA, hek-293 cells, fetal tissue research, abortion used in vaccine, organ harvesting, conscience vaccines abortion \\
\bottomrule
\end{tabular}
\caption{The final set of 48 search queries spanning 8 misinformation topics employed in our audit study. For each topic, we utilized 6 search queries associated with the topic.}
\label{tab:full-query-set}
\end{table*}

\section{Curating and Validating Search Queries}\label{sec:appendix_val1}

To conduct all our validation experiments, we curated a set of 18 search queries\footnote{Such queries were: Day in the life, City vlogs, Travel, Hiking, University, Food, School, Places nearby, Things to do nearby, news near me, Hospital, Bank, Government, Politics, Businesses, Stores, Groceries, and Bars.} based prior work \cite{10.1145/2815675.2815714} that we thought were likely to elicit geolocation-based personalization. We also incorporated 5 random search queries curated from Section \ref{sec:compile}, giving us 23 search queries total. To validate that the search queries elicit geolocation-based personalization, we conducted a validation experiment in Seattle, Washington, US, and Cape Town, Western Cape, SA, placing a twin bot in each geolocation. We ran the experiment from July 17th to 19th, 2022 at three evenly distributed times of 00:00 UTC, 08:00 UTC, and 16:00 UTC. Using the \texttt{GBP} metric from Appendix \ref{sec:personalization_metric}, the average \texttt{GBP} was 0.18, indicating that 18\% of the search results differed between Cape Town, South Africa, and Seattle, Washington, even after accounting for noise. This validated that our curated search queries elicited geolocation-based personalization on YouTube. We used the search queries in the subsequent validation experiments

\section{Validating Geospoofing Method}\label{sec:appendix_val2}

With a validated set of search queries, we tested whether YouTube personalizes search results based on GPS coordinates provided rather than IP addresses by comparing the search results of bots with different IP addresses but with the same provided GPS coordinates. To override the bot's current geolocation with our provided GPS coordinates, we used Chrome DevTool \cite{chrome_devtools} to integrate the geolocation emulation feature with our Selenium bots. We created twin bots in Cape Town (SA), Paris (France), and North California (US) regions on AWS, giving us 6 bots total. We provided all bots with the same GPS coordinate in Paris, France. As such, twin bots in varying countries should have different IP geolocations, but the same geospoofed GPS coordinates. If YouTube used GPS coordinates to personalize SERPs, we would see no difference in SERPs after accounting for noise because the bots have the same GPS coordinates, resulting in a \texttt{GBP} metric value of 0. However, if YouTube used IP addresses, the \texttt{GBP} metric value would be greater than 0. From July 25th to July 27th, 2022, we ran the validation experiment at three evenly distributed times 00:00 UTC, 08:00 UTC, and 16:00 UTC. The average \texttt{GBP} value between North California and Paris was 0.15, while the average \texttt{GBP} value between Cape Town and Paris was 0.11. Our results indicate that YouTube uses IP addresses over the provided geospoofed GPS coordinates to personalize their SERPs. Given this result, we shifted towards designing our experiment using proxies. 

\section{Validating Proxies for Accurate IP Location}\label{sec:appendix_val3}

After obtaining unsuccessful validation results from the geospoofing method, we conducted a third experiment to validate that \textit{IPRoyal} provides the bots with accurate IP geolocations. We created 2 sets of twin bots in Cape Town, South Africa on AWS, giving us 4 bots total. A set of twin bots would search regularly, while the other set of twin bots would search using the \textit{IPRoyal} proxies set to Cape Town, South Africa. Since we are comparing bots in the same geolocation, we should ideally see no geolocation-based personalization. Additionally, we used IP2Location to keep track of IP addresses \cite{ip2location}, enabling us to validate whether the \textit{IPRoyal} proxies are associated with our desired IP geolocation. Since the 2 sets of twin bots were at the same IP geolocation, our \texttt{GBP} metric value should be close to 0. From December 22nd to December 24th, 2022, we ran the validation experiment at two evenly distributed times of 00:00 UTC and 12:00 UTC. The average geolocation-based personalization value was -0.02,\footnote{Note that, our formalized metric from Appendix \ref{sec:personalization_metric} may result in a slight negative value since we take the more restrictive noise to establish our baseline noise. Regardless, any slightly negative value indicates that the differences between the treatments did not exceed the baseline noise.} validating that the bots had no geolocation-based personalization. Additionally, our IP2Location data also validates the consistency and accuracy of \textit{IPRoyal} in providing proxies with valid IP geolocations in Cape Town. 

\begin{table*}[ht]
\centering
\small
\begin{tabular}{p{2.3cm}p{2.6cm}p{8cm}}
\toprule
\textbf{Proxy Provider} & \textbf{Price (\$)} & \textbf{Proxy Geolocations in South Africa (City)} \\ 
\midrule
Bright Data & \$30 / GB of traffic & Benoni, Bloemfontein, Boksburg, Cape Town, Centurion, Ceres, Durban, Johannesburg, Pietermaritzburg, Polokwane, Port Elizabeth, Pretoria, Queenstown, Rabieridge, Tembisa, Tzaneen, Vanderbijlpark \\\midrule
Smart Proxy & \$10 / GB of traffic & N/A \\\midrule
OxyLab & \$15 / GB of traffic & Cape Town \\\midrule
IPRoyal & \$2--4 / GB of traffic & Cape Town, Johannesburg, Durban, Pretoria \\\midrule
NordVPN & \$11.99 per month & N/A \\\midrule
ExpressVPN & \$12.95 per month & N/A \\
\bottomrule
\end{tabular}
\caption{Proxy providers with services available in South Africa. We list each proxy provider's price (per gigabyte of traffic data or monthly subscription) and the proxy geolocations available in South Africa. Some proxy providers do not have a dedicated pool of proxies in specific geolocations/cities in South Africa, which we denoted as ``N/A'' in the column ``Proxy Geolocations in South Africa (City).'' Except for \textit{IPRoyal}, all proxy providers were expensive (e.g., $>$\$10 / GB of traffic) or did not have proxies available in our selected geolocations in SA.}
\label{tab:proxy}
\end{table*}

\section{Amazon Mechanical Turk Job}\label{sec:mturk}

This section describes how we obtained annotations from Amazon Mechanical Turk (AMT) workers. We implemented a rigorous screening process for AMT workers to ensure high-quality annotations. In the following sections, we briefly provide details on our screening process and the annotation task. 

\subsection{Training and Screening Workers}
To get high-quality annotations, we trained and screened AMT workers by adding two qualification requirements. First, to find workers with a proven track record of high-quality work, we required AMT workers to have at least 1000 approved tasks with at least a 99\% approval rating on the AMT platform. Second, we required the AMT workers to get a full score of 100 on our Qualification Test. We introduced the test to train and ensure that AMT workers attempting our annotation task understood our annotation scheme well. The Qualification Test consisted of 5 questions, with one eligibility question asking them to confirm whether they are affiliated with the authors' university. The other four questions asked AMT workers to annotate YouTube videos whose annotation labels were known in advance---these videos were already annotated by the authors (see Figure \ref{fig:annotation-task} for a sample Qualification Test question). To ensure that the AMT workers understood the task and annotation scheme, we gave detailed instructions and described each annotation value in detail with various examples of YouTube videos in the Qualification Test (Figures \ref{fig:qualification-example}, \ref{fig:qualification-test}, \ref{fig:qualification-instruction}). For each video example, we provided the video metadata (e.g. video title, embedded Video URL) and explained why a particular annotation label was assigned to the video (see Figure \ref{fig:qualification-example}).

We took two steps to ensure the instructions, test questions, and overall annotation task were clear and comprehensive. First, we posted the Qualification Test and the annotation task on r/mturk,\footnote{https://www.reddit.com/r/mturk/} a subreddit community of AMT workers, and Turker Nation, an unofficial Slack channel of AMT workers. Second, we conducted a pilot run by posting 13 annotation tasks and the aforementioned screening requirements. Our task received positive feedback from the AMT community, which indicates acknowledgment of our task design. One respondent reported \emph{``Seems interesting. Hope to get to work on it when it's ready,''} while another stated \emph{``The test was fairly easy and constructed well.''} After obtaining positive feedback from the AMT communities and a successful pilot run, we released our AMT annotation task titled ``YouTube Video Labeling (COVID-19).'' We paid the AMT workers the United States federal minimum wage (\$7.25 per hour). When refining our annotation heuristics in \S \ref{annotation_dev}, we kept track of the time it took the external researchers to annotate the 13 videos, which ranged from 26 seconds to 42 minutes in video duration. We found that it took, on average, 1.5 minutes to annotate a video. However, we erred on the side of overestimation and used 2 minutes as the baseline to annotate a video. We prorated the federal minimum wage using the 2-minute annotation time, thus paying each annotator \$0.24 per annotation. Our compensation is much higher than prior works \cite{papadamou, 10.1145/3555618, liaw2023younicon}; for example, \citet{papadamou} paid \$0.03 per annotation for labeling YouTube videos for pseudoscientific content.

\subsection{YouTube Annotation Task}

In our annotation task, we required each AMT worker to assign an annotation label to the video and provide a short rationale behind selecting the label. We required a minimum of 10 characters for our rationale. We released 1,988 videos on AMT in batches of 15 at a time. Three different AMT workers annotated each video. Throughout the annotation process, we sampled videos in each batch to ensure quality work and reasonable rationales from the AMT workers.\footnote{We found that two AMT workers submitted identical annotations and rationale across the tasks; we rejected their work.} Using the collected annotations, we calculated the Fleiss' Kappa Score ($\kappa$) to assess the annotators' agreement. We obtained $\kappa = 0.62$, which is considered ``substantial agreement.'' Our $\kappa$ value of 0.62 is a much higher agreement score compared to a close prior work; \citet{papadamou} asked three AMT workers to annotate YouTube videos for pseudoscientific content, including COVID-19 conspiracy theories, using 3 labels, resulting in a Fleiss' $\kappa$ score of 0.14. 

To determine the final label of the YouTube video, we chose the annotation label with at least a majority agreement among the workers. We arrived at a final label for 1,899 videos through a majority agreement among the workers. For the remaining 89 videos, all three AMT worker responses differed. The first author, as the expert, annotated these YouTube videos to arrive at the final annotation label. Figure \ref{fig:annotation-ui} shows the interface of our AMT annotation task.

\section{Classifier Training Details}\label{sec:classifier-training-details}
This section describes the selected input features and training procedures of the chosen classification models.

\subsection{Feature Descriptions.} We considered the following input features for our classifier.

\noindent\textbf{Video Title: }The title of the video.\\
\noindent\textbf{Video Description: }A small description regarding the content of the video.\\
\noindent\textbf{Video Transcript: }Transcript contains the video's actual content, which is either subtitles uploaded by the content creator or auto-generated by YouTube. This feature often contains the main themes discussed by the creator/uploader of the video.\\
\noindent\textbf{Video Tags: }As mentioned in \S\ref{sec:methods}, video tags are descriptive keywords representing how content creators want their videos to be discovered. The content creators can specify relevant tags to associate with their videos during the upload process. \\
\noindent\textbf{Video Comments: }\citet{medina-serrano-etal-2020-nlp} found that YouTube comments was a highly predictive feature for detecting COVID-19 misinformation in YouTube videos. Thus, we selected the top 100 comments associated with each video. \\

\subsection{Training Procedures} 
Here, we outline the various training procedures and hyperparameters employed to train various classification models. 

\begin{table}[t!]
\centering
\small
\begin{tabular}{@{}lp{5.5cm}@{}}
\toprule
\textbf{Model} & \textbf{Hyperparameter Search Space} \\
\midrule
SVM & 
  \begin{tabular}{@{}l@{\hspace{0.2em}}l@{}}
    C: & [0.1, 1, 10, 100, 1000] \\
    $\gamma$: & [1, 0.1, 0.01, 0.001, 0.0001] \\
    Kernel: & [rbf]
  \end{tabular} \\
\midrule
XGBoost & 
  \begin{tabular}{@{}l@{\hspace{0.2em}}l@{}}
    max\_depth: & [2, 3, 4, 5, 6, 7, 8, 9, 10] \\
    n\_estimators: & [60, 100, 140, 180, 220] \\
    learning\_rate: & [0.5, 0.1, 0.01, 0.05, 0.001]
  \end{tabular} \\   
\bottomrule
\end{tabular}
\caption{Hyperparameter search space for each ML model. Grid search with 5-fold cross-validation was used to find optimal hyperparameters with highest average accuracy.}
\label{tab:gridsearch}
\end{table}

\subsubsection{Support Vector Machines and XGBoost.}
For traditional machine learning algorithms such as SVMs and XGBoost, we applied standard preprocessing procedures on the selected features, such as removing stopwords and whitespace. To create the feature vectors, we tried out 4 different vectorizers: count, TF-IDF, FastText \cite{bojanowski2017enriching}, and Word2Vec \cite{mikolov2013efficient}. To find the best set of hyperparameters, we employed a grid search strategy with 5-fold cross-validation on our training dataset, exploring each model's defined hyperparameter search space (see Table \ref{tab:gridsearch}) and recording the set of hyperparameters that resulted in the highest average cross-validation accuracy. We experimented with various combinations of models, vectorizers, and feature sets. 

\subsubsection{DeBerta-v3-base and DeBerta-v3-large.} From training our SVM and XGBoost models, we found that video comments as an input feature resulted in poor performance for our task (see Appendix \ref{sec:classifier-performance}). Thus, we concatenated the video's title, description, transcript, and tags as our input text. We intentionally excluded including the video comment as a feature due to their poor performance in predicting COVID-19 misinformation in videos. The concatenated input text was then truncated, retaining the first 1,024 tokens. In addition to the 10\% held-out test set, we set aside 10\% of the ground-truth dataset to serve as our validation set, leaving 80\% of the dataset as our training set. To train the model, we used the Adam optimizer and cross-entropy loss as our loss function. To avoid overfitting, we employed early stopping, in which we trained the model until the validation loss no longer improved over several iterations. We conducted extensive hyperparameter tuning and arrived at each model's final set of hyperparameters. For the DeBerta-v3-base, we employed a training batch size of 8, a learning rate of 1e-5, and a weight decay of 1e-3. For the DeBerta-v3-large, we employed a training batch size of 4, a learning rate of 5e-6, and a weight decay of 1e-4. These models were trained on a single NVIDIA A40 GPU.

\subsubsection{GPT-4 Turbo.}
We describe the prompt design considerations and tuning process to evaluate GPT-4 Turbo's performance in detecting COVID-19 misinformation based on YouTube video metadata. Since we employed an OpenAI's GPT-4 model, we designed our prompts based on OpenAI's recommendations on prompt-engineering \cite{openai_promptengineering} and prior work \cite{10.1145/3639476.3639777}.

First, prior research showed that models demonstrate improved performance when compelled to reason and justify their decisions \cite{openai_promptengineering, dammu-etal-2024-uncultured}. Therefore, we mandated GPT-4 to provide direct excerpts from the YouTube video metadata and concise justifications regarding their selected label. 

Second, \citet{openai_promptengineering} recommended that asking the model to adopt a ``persona'' in their system can lead to better results from LLMs. Thus, for our task, we prompted GPT-4 to adopt a persona as a ``public health expert'' with comprehensive knowledge of the COVID-19 pandemic and its misinformation. 

Third, we experimented with both zero-shot prompting and few-shot prompting. Zero-shot prompts involve presenting the task to the LLM, only including the label descriptions without accompanying examples or training.
Meanwhile, few-shot prompts involve including task-specific examples within the prompt, enabling the pre-trained LLM to condition on the illustrative examples rather than updating its weights. We provided five few-shot examples, each containing video metadata, an assigned label, and reasoning for the label. These few-shot examples were the same as the ones provided to AMT workers on their annotation tasks. 

Fourth, we experimented with varying temperature levels to find our task's most ideal parameter setting. Ranging from 0 to 2, the temperature parameter influences how models generate text. When utilizing lower values for temperature, such as 0, the generated text becomes deterministic, selecting more consistent and coherent outputs.Previous works \cite{10.1145/3639476.3639777, park-etal-2024-valuescope} also experimented with various temperature values, such as 0.2 and 0.7, for their text annotation tasks and found that lower temperature values (e.g. 0.2) resulted in the best performance. Likewise, we selected relatively lower temperature values: 0, 0.2, and 0.7 for our experiments. We used default settings for other parameters. See Table \ref{tab:zero-shot-prompt} for our zero-shot prompt and Table \ref{tab:few-shot-prompt} for our few-shot prompt.

Given the poor performance of video comments as an input feature when training SVM and XGBoost for our task (see Appendix \ref{sec:classifier-performance}), we supplied the model with video metadata, such as the video title, description, transcript, and tags, within the prompts. To manage the cost of GPT-4's API based on token usage, we took measures to efficiently control expenses by truncating the video metadata. We truncated the video metadata in our prompt, restricting it to the first 500 tokens per video metadata supplied. This also included the video metadata included in the few-shot examples. To check whether truncating the video metadata impacts GPT-4's performance, we also evaluated how GPT-4 performs when given prompts with full metadata vs. prompts with truncated metadata. 

\section{Classifier Performance Results}\label{sec:classifier-performance}
This section describes the classifier performance results and analysis to understand which input features contribute most to detecting COVID-19 misinformation in YouTube videos. 

\subsection{Performance Results}
In total, we trained and evaluated a total of 62 different trained classifiers. To assess the models consistently, we evaluated the model over the same 10\% held-out test set (273 videos) from our ground-truth dataset, providing insights into the model's generalizability and unbiased performance over unseen data. We employed three metrics to capture performance: accuracy, weighted F1-score, and macro F1-score. Table \ref{tab:classifier-full-results} reports the performance of all classifiers. Our best-performing model was DeBerta-v3-large (Index 54 in Table \ref{tab:classifier-full-results}), which scored 0.85 across accuracy, weighted F1-score, and macro F1-score. Among traditional machine learning classifiers, we found that SVMs performed the best when using titles, descriptions, transcripts, and tags as the feature set with a TFIDF vectorizer (Index 33 of Table \ref{tab:classifier-full-results}), scoring 0.78 across the performance metrics. 

Notably, GPT-4 Turbo performed the best when provided with a zero-shot prompt comprising truncated title, description, transcript, and tags as metadata, along with a temperature of 0.2 (Index 57 of Table \ref{tab:classifier-full-results}), achieving 0.79 across the performance metrics. Despite not being given any few-shot examples or extensively fine-tuned on the task, zero-shot GPT-4 outperformed the best-performing traditional machine-learning algorithms (i.e., the aforementioned SVM model). In the same setting (e.g. zero-shot, temperature=0), we also evaluated prompts where we provided truncated video metadata to the first 500 tokens (Index 56) vs. complete video metadata in the prompt (Index 55). We found that truncating the video metadata did not negatively impact GPT-4's performance; in fact, our results suggest that prompts with truncated video metadata improved GPT-4's performance (0.78 across performance metrics) compared to that of complete video metadata (0.76 across performance metrics). This may be attributed to the complex and long prompts when using full video metadata. Thus, the improved performance when using truncated metadata aligns with \citet{openai_promptengineering}'s recommendation of utilizing shorter, concise details to simplify the prompt. 

\subsubsection{Which video features are important in detecting COVID-19 misinformation?}

To understand which input feature contributes most to the classification of COVID-19 misinformation videos, we systematically evaluated each feature individually and in combination with other features. Due to the computational and financial costs of employing GPT-4 Turbo and DeBerta-v3, we opted for SVM with FastText as our vectorizer. The training procedure for SVMs remained consistent with the approach outlined earlier. 

Performance results for each individual feature and combination of features are presented in Index 0-13 of Table \ref{tab:classifier-full-results}. Among the individual features, we observed that video titles yielded the highest performance (0.64 accuracy), suggesting that they are an informative input feature. Similarly, descriptions (0.6 accuracy) and transcripts (0.64 accuracy) demonstrated comparable performance. In contrast to the findings of \citet{medina-serrano-etal-2020-nlp}, where they identified video comments as highly predictive in detecting COVID-19 misinformation in YouTube videos, our results revealed video comment as the least effective input feature, achieving an accuracy of only 0.49. Combining title, description, and transcript (Index 9) or title, description, transcript, and tags (Index 12) yielded the best performance of 0.69 accuracy. This further indicates that video comments are not informative for our classification task.

\begin{table}[t!]
\centering
\small
\begin{tabular}{c|c|c|c|c}  
\toprule
\textbf{Top-N Videos} & \textbf{\textit{p}} & \textbf{\textit{r}} & \textbf{\textit{U}} & \textbf{Mean Rank Diff.} \\ 
\midrule
10 & *** & 0.49 & 191 & SA $>$ US \\\midrule
20 & * & 0.30 & 293 & SA $>$ US \\\midrule
30 & - & - & 429 & - \\\midrule
40 & * & 0.31 & 285 & US $>$ SA \\\midrule
50 & ** & 0.36 & 256 & US $>$ SA \\
\bottomrule
\end{tabular}
\caption{Results of the Mann-Whitney U Tests to compare the misinformation bias scores between US and SA at each level considering the top-N videos in the SERPs. For each test, we denote the p-value (\textit{p}), Mann-Whitney effect size (\textit{r}), U-value (\textit{U}), and the mean rank difference. For example, when considering the top-10 videos, SA $>$ US indicates that bots in SA received more misinformative videos in the top-10 search results than bots in the US. Note that: *$P < 0.05$; **$P < 0.01$; ***$P <0.001$. Results with no significance was denoted with -.}
\label{tab:top10-50-mann-whitney}
\end{table}

\section{Significance Testing Details}\label{sec:reproducible}
For replicability, we outline the inputs, variables, and numbers involved in the Mann-Whitney U and Kruskal-Wallis H tests.

\subsection{Mann-Whitney U Test for Top-N Videos in SERPs.} As shown in Figure \ref{fig:top10-50-distribution} and Table \ref{tab:top10-50-mann-whitney}, we performed a Mann-Whitney U Test to compare the misinformation bias scores between US and SA geolocations at each level considering the top-N videos in the SERPs. The independent variable was the geolocations in the US vs. SA, while the dependent variable was the misinformation bias scores. The misinformation bias score was calculated by considering the top 10 to top 50 videos in the SERPs. The misinformation bias scores were averaged across all search queries, search filters, and the twin bots associated with each geolocation. Each geolocation resulted in 10 averaged misinformation bias scores, each score value corresponding to a particular day during our experimental run. Given that we had 3 geolocations per country, we combined the averaged scores for each country, resulting in 30 scores for the US and 30 scores for SA. We conducted the Mann-Whitney U Test to compare the misinformation bias scores between the US and SA at each level of top-N videos in the SERPs. We record the p-value, Mann-Whitney effect size \textit{r}, U-value, and mean rank difference in Table \ref{tab:top10-50-mann-whitney}.

\begin{figure}[t!]
  \centering
  \includegraphics[width=1\linewidth]{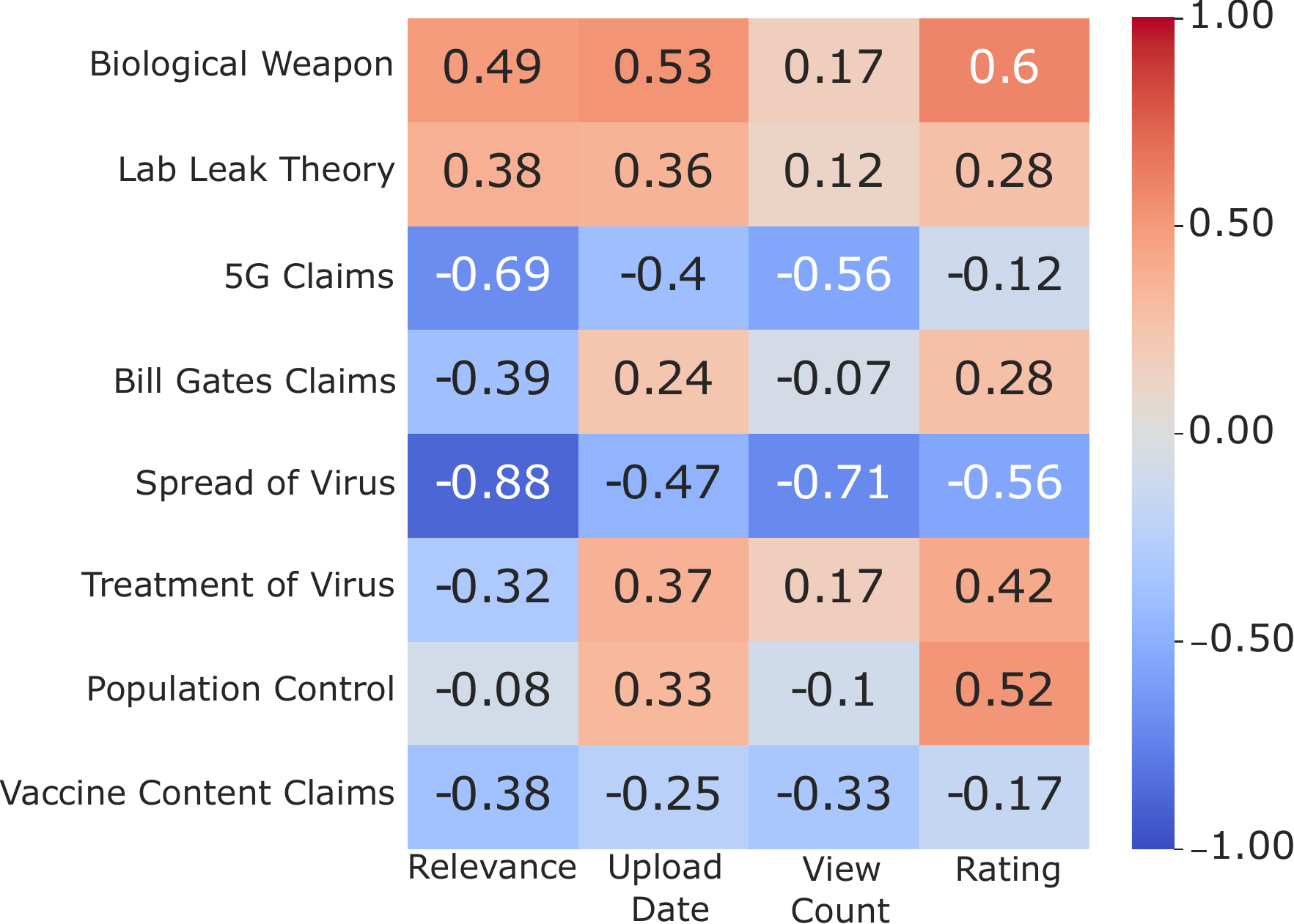}
  \caption{The mean misinformation bias scores for all 8 topics across 4 search filters based on all SERPs collected in the US and SA. Note that the bias scores were calculated considering the top-10 search results.}
  \label{figure:heatmap_overall}
\end{figure}

\subsection{Kruskal-Wallis H Test for Within-Country Comparison}
In the Results section (\S \ref{sec:results}), we performed Kruskal-Wallis H Test to compare the misinformation bias scores within the geolocations in a given country. The independent variable was the 3 selected geolocations within a country, while the dependent variable was the misinformation bias scores. The misinformation bias scores were calculated by considering the top-10 search results in the SERPs. For each geolocation, we calculated the average misinformation bias scores across the search queries, search filters, and twin bots associated with that geolocation. This resulted in 10 mean misinformation bias scores per geolocation, where each score value corresponded to a particular experimental run. To assess the differences among the geolocations within the country, we conducted the Kruskal-Wallis H Test using the mean misinformation bias scores from the three geolocations within a country, resulting in a total of 30 samples (10 samples each from three geolocations). We also computed the Kruskal-Wallis Test effect size \cite{kruskal-effect}, denoted as $\eta^2$. Following a significant result from the Kruskal-Wallis H Test, we conducted a post-hoc Conover-Iman Test with Bonferroni adjustment using the same inputs to examine the pairwise differences between geolocation. We repeated this procedure for each country.

\begin{figure}[t!]
  \centering
  \includegraphics[width=1\linewidth]{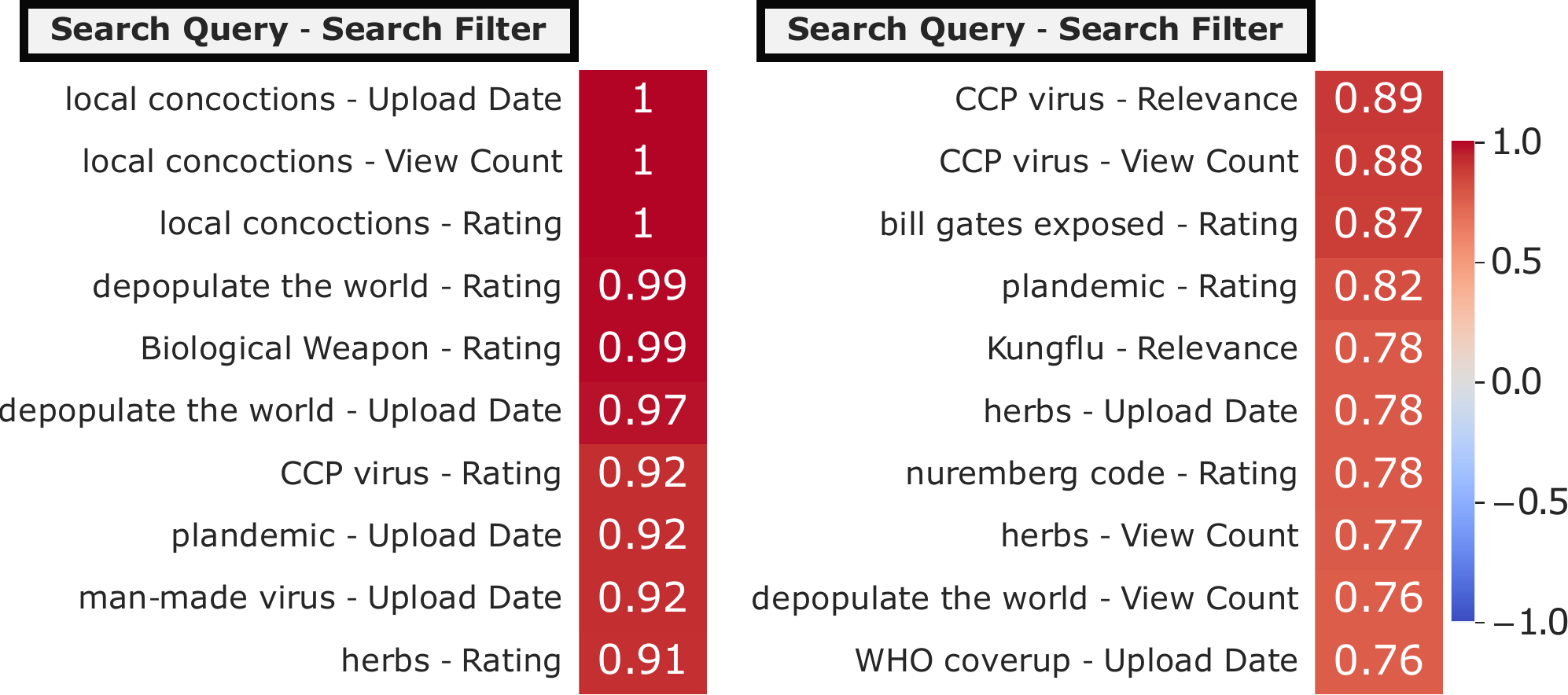}
  \caption{Top 20 search query-filter combinations when sorted by the mean misinformation bias score. These query-filter combinations are the most problematic ones, containing the highest amount of misinformation in the search results.}
  \label{figure:search-queries}
\end{figure}

\subsection{Mann-Whitney U Test for Topic-Wise Comparisons}
In Figure \ref{fig:topic_diff}, we performed Mann-Whitney U Tests to compare the misinformation bias scores between the US and SA geolocations for each of the 8 topics. The independent variable was the geolocations in the US and SA, while the dependent variable was the misinformation bias score, considering the top-10 search results in the SERPs. For a particular topic, we averaged the misinformation bias scores across the topic's constituting search queries, search filters, and twin bots associated with each geolocation. Each geolocation resulted in 10 averaged misinformation bias scores for a particular topic; each score value corresponding to a particular day during our experimental run. Given that we had 3 geolocations per country, we combined the averaged scores for each country, resulting in 30 scores for the US and 30 scores for SA for the particular topic. Subsequently, we conducted a Mann-Whitney U Test for each topic to compare the misinformation bias scores between the US and SA. We record the p-value, Mann-Whitney effect size \textit{r}, U-value, and mean rank difference in Figure \ref{fig:topic_diff}.

\subsection{Mann-Whitney U Test for Filter-Wise Comparisons}
In Figure \ref{fig:filter_diff}, we performed Mann-Whitney U Tests to compare the misinformation bias scores between the US and SA geolocations for each of the 4 search filters. The independent variable was the geolocations in the US and SA, while the dependent variable was the misinformation bias score, considering the top-10 search results in the SERPs. For a particular search filter, we averaged the misinformation bias scores across the search queries and twin bots associated with each geolocation. Each geolocation resulted in 10 averaged misinformation bias scores for a particular filter, each score value corresponding to a particular day during our experimental run. Given that we had 3 geolocations per country, we combined the averaged scores for each country, resulting in 30 scores for the US and 30 scores for SA for a particular filter. Subsequently, we conducted a Mann-Whitney U Test for each filter to compare the misinformation bias scores between the US and SA. We record the p-value, Mann-Whitney effect size \textit{r}, U-value, and mean rank difference in Figure \ref{fig:filter_diff}.

\section{Additional Analysis}\label{sec:additional-analysis}
In this section, we analyze the misinformation bias in search queries and temporal trends of the misinformation bias scores between bots in the US and SA. We also provide Figure \ref{figure:heatmap_overall}, which shows the heatmap of the mean misinformation bias scores for all 8 topics across 4 filters considering all SERPs collected in the US and SA.

\subsection{Misinformation Bias in Search Queries}
Figure \ref{figure:search-queries} shows the top 20 search queries and filter combinations with the highest misinformation bias scores. Surprisingly, 10 search query-filter combinations have a very high misinformation bias score ($>$ 0.9). The search query ``local concoctions'' yields a bias score 1 for three filter types, indicating that the search results are completely plagued with misinformative content. Most of the search queries in Figure \ref{figure:search-queries} have a negative connotation, i.e. the search queries themselves have a bias (e.g. the queries ``plandemic'' and ``depopulate the world'' indicates an intent to search for misinformative, conspiratorial content). This observation reveals that if you search for misinformation, you will get high amounts of misinformative search results. This indicates how current search engines work; they curate and recommend content without necessarily considering the content's veracity. The most troublesome observation is the presence of high misinformation bias for misleading, xenophobic queries, ``CCP virus'' (0.88-0.92) and ``Kungflu'' (0.78). This suggests that YouTube frequently returns highly misinformative search results for such misleading terms, potentially perpetuating their usage.

\subsection{Temporal Trends}
\begin{figure}[!t]
  \centering
  \includegraphics[width=1\linewidth]{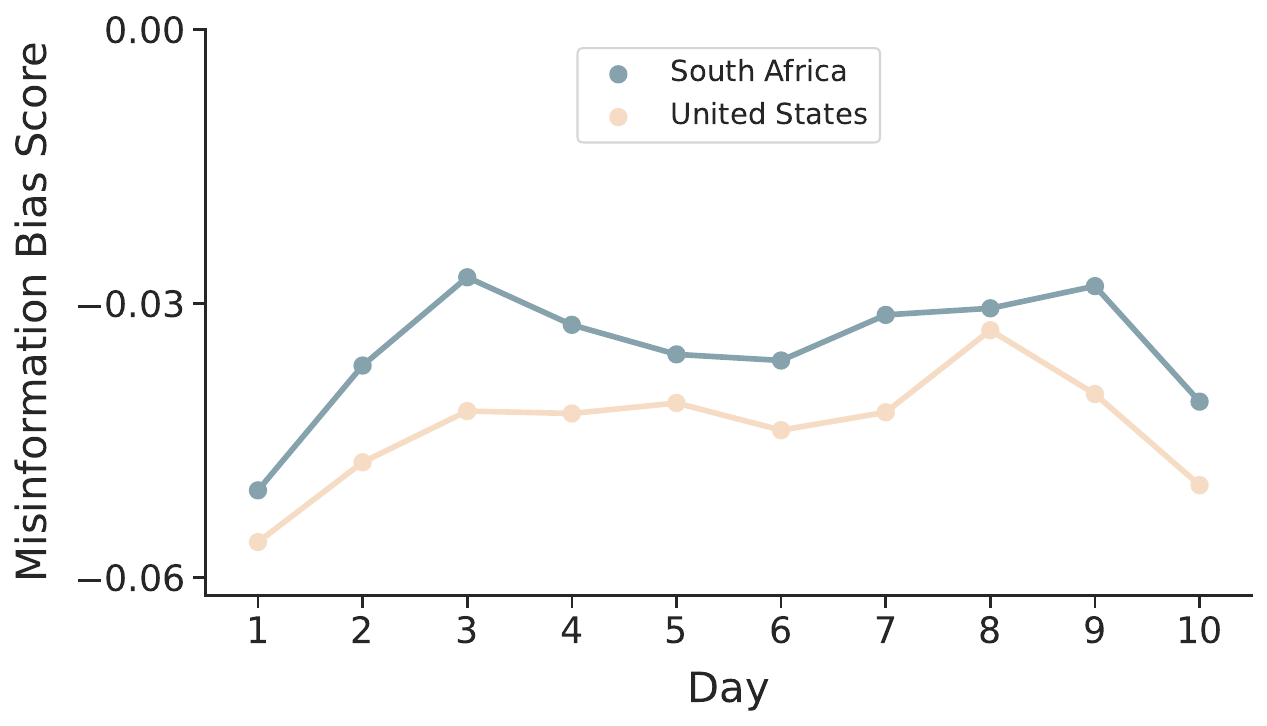}
  \caption{Mean misinformation bias scores of SERPs between bots in the US and SA for each day of the experiment run. Over the 10 days, bots in the SA consistently maintained a higher bias score than bots in the US.}
  \label{figure:temporal_trends}
\end{figure}

\begin{figure*}[!ht]
  \centering
  \includegraphics[width=0.8\textwidth]{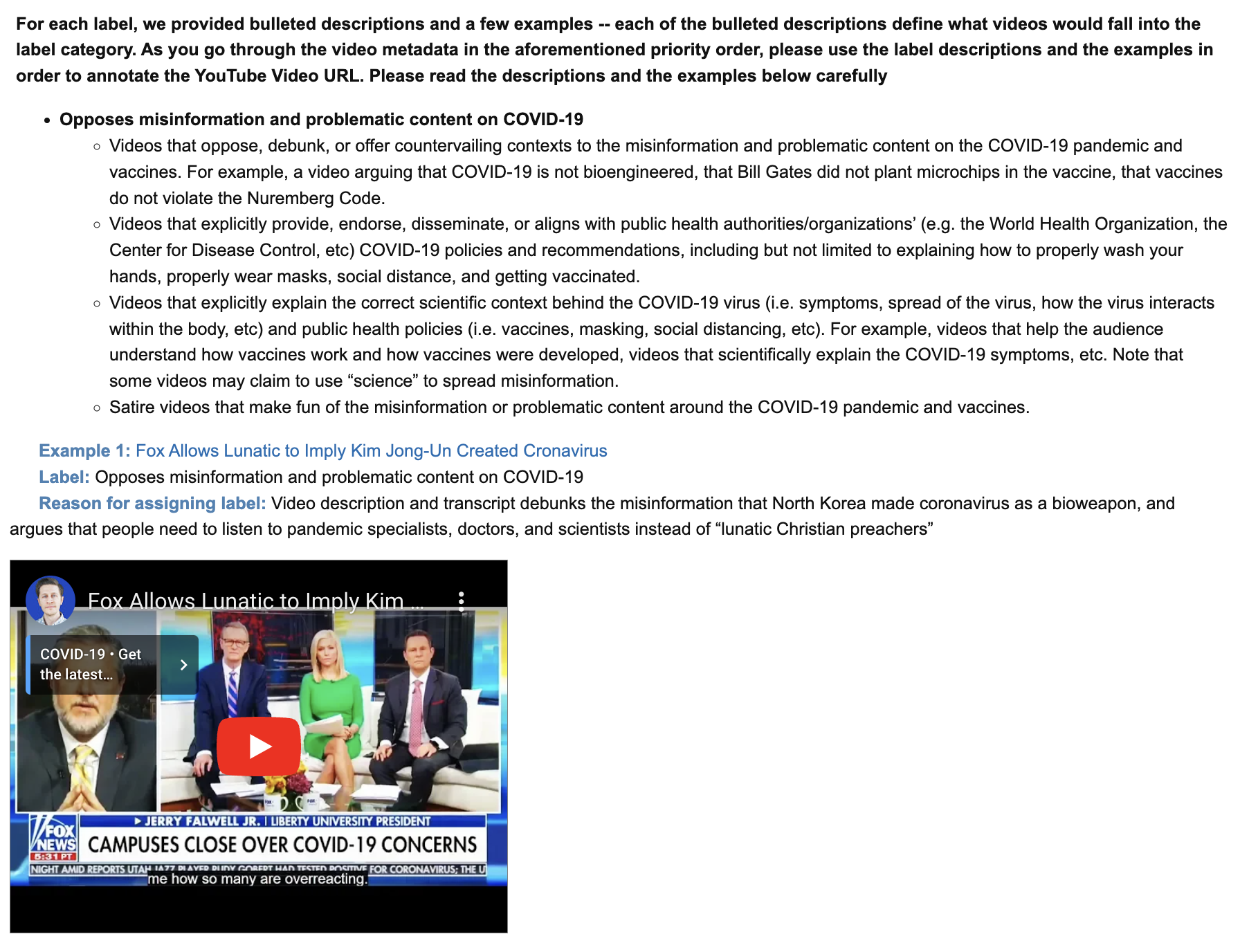}
  \caption{For each annotation label, we provided bulleted descriptions and a few examples, including the reasoning behind assigning the label. These annotation label descriptions were provided in both the Qualification Test and the actual annotation task.}
  \label{fig:qualification-example}
\end{figure*}

\begin{figure*}[!ht]
  \centering
  \includegraphics[width=\textwidth]{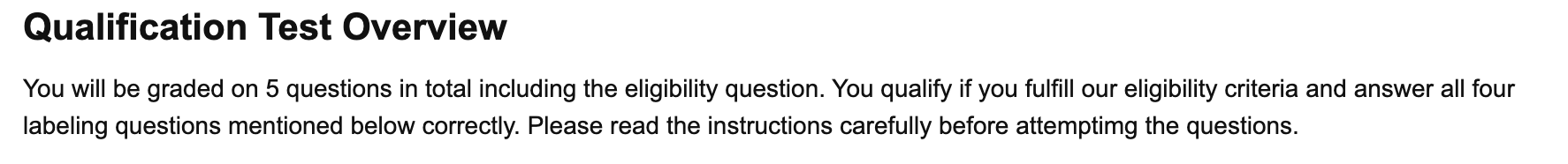}
  \caption{Figure illustrating our Qualification Test instructions. Our test included 5 questions, of which four were labeling questions and one was an eligibility question that required the addition of the authors' university. A full score of 100 was required to qualify for the test.}
  \label{fig:qualification-test}
\end{figure*}

\begin{figure*}[!ht]
  \centering
  \includegraphics[width=0.8\textwidth]{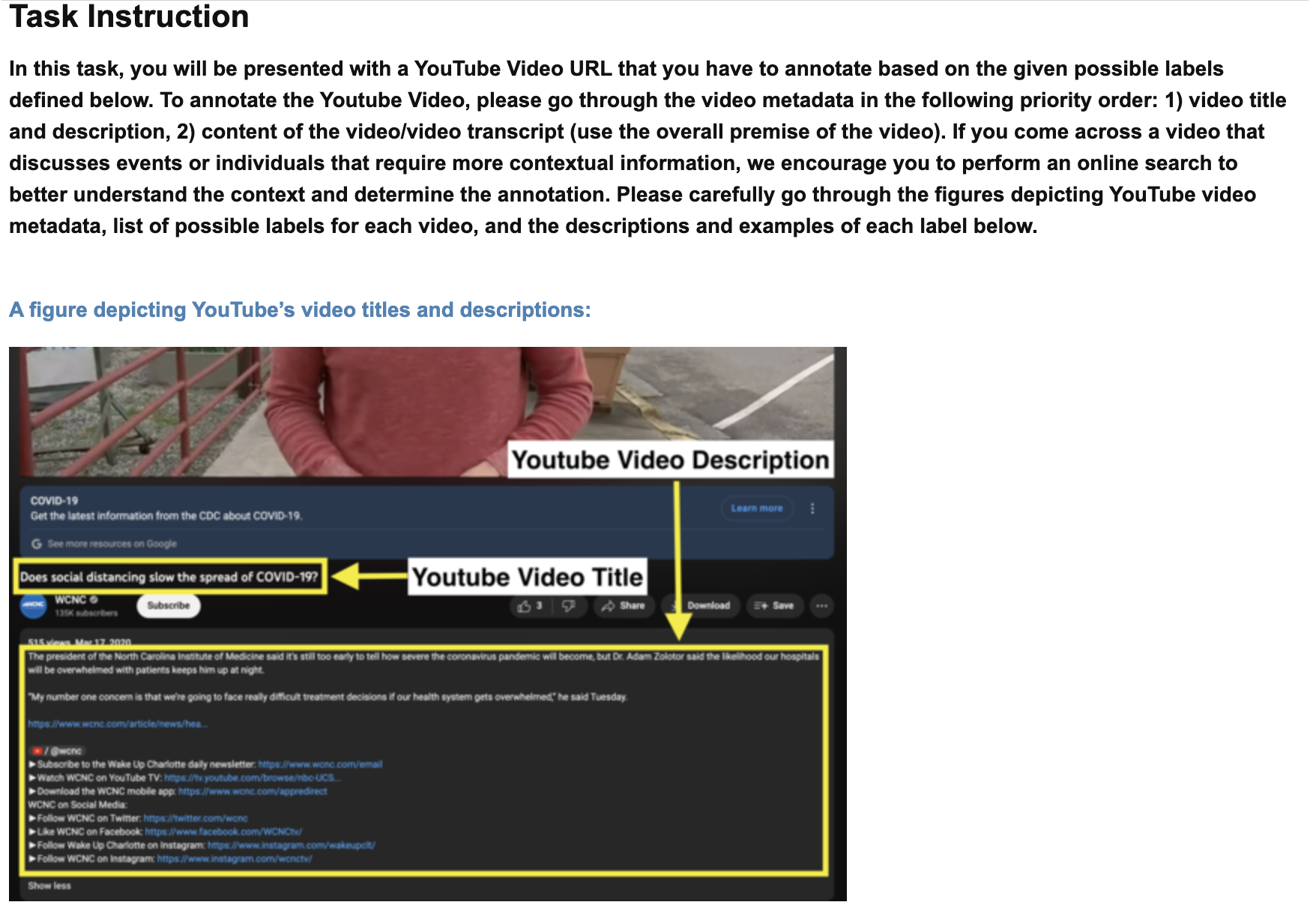}
  \caption{Task instruction in the Qualification Test explaining how to annotate YouTube videos for our task. The same instructions were provided in the actual annotation task.}
  \label{fig:qualification-instruction}
\end{figure*}

\begin{figure*}[!ht]
  \centering
  \includegraphics[width=0.8\linewidth]{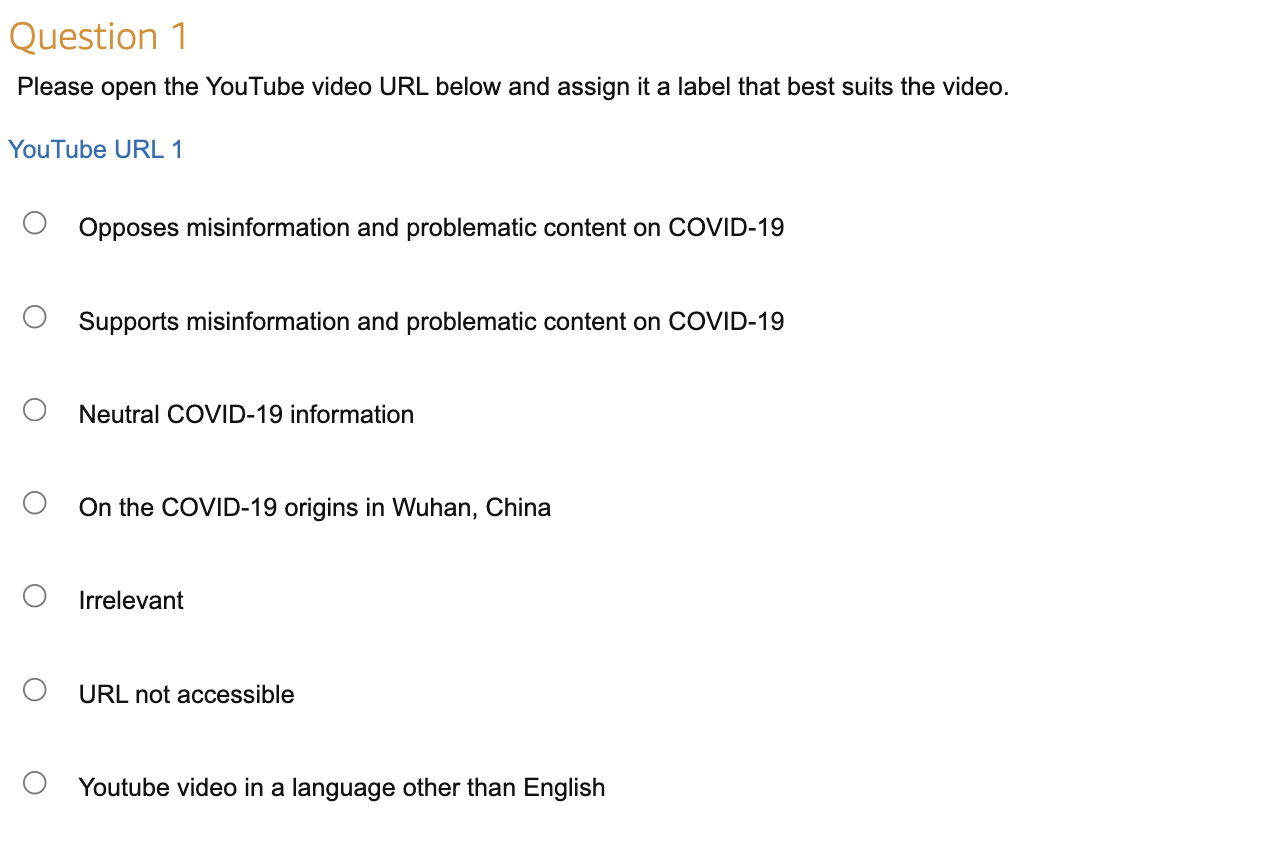}
  \caption{Example of the Qualification Test question.}
  \label{fig:annotation-task}
\end{figure*}

\begin{figure*}[!ht]
  \centering
  \includegraphics[width=0.8\textwidth]{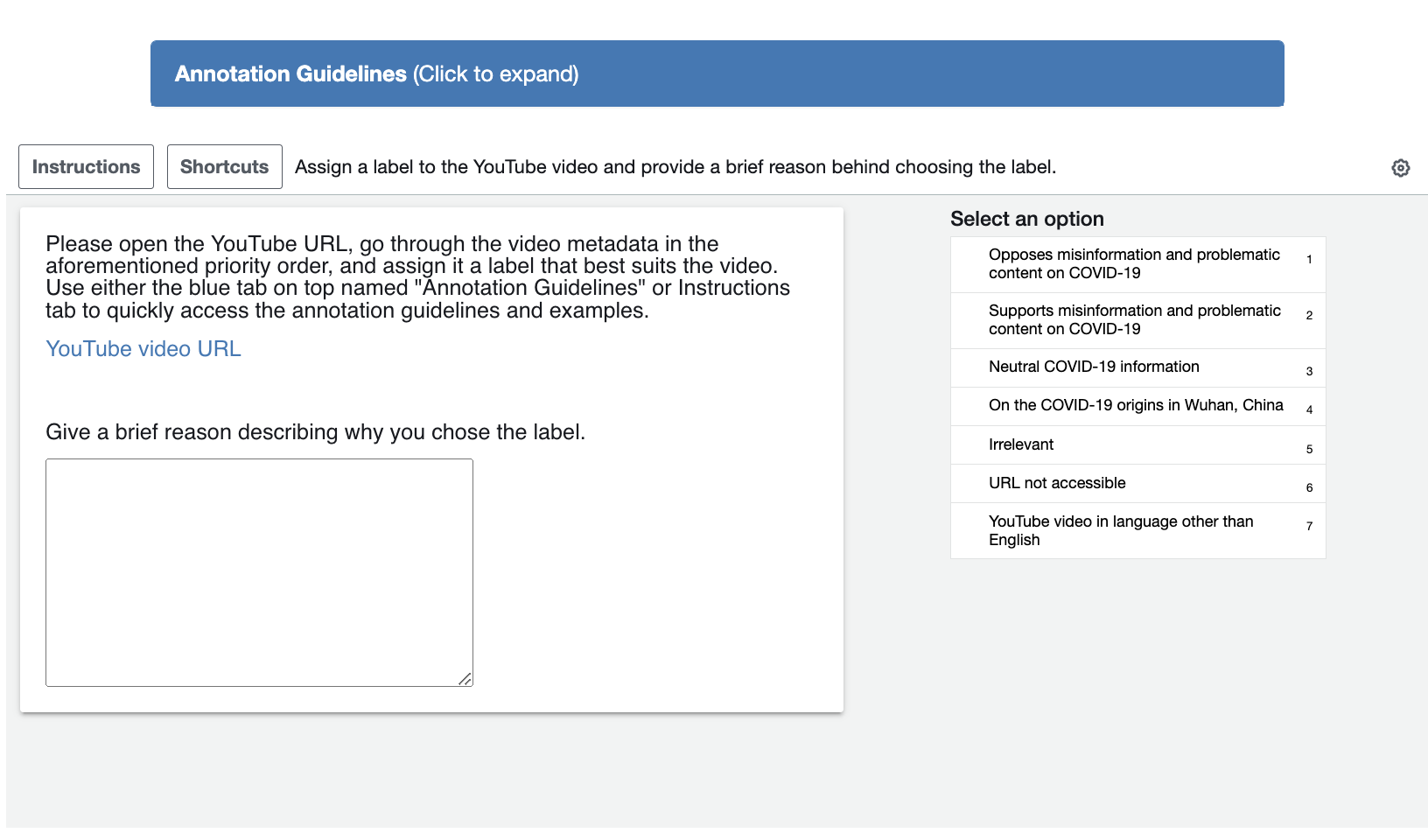}
  \caption{Interface of our YouTube video annotation task on Amazon Mechanical Turk.}
  \label{fig:annotation-ui}
\end{figure*}

Figure \ref{figure:temporal_trends} shows the mean misinformation bias scores of SERPs between bots in the US vs. SA across the 10 experimental days. Each point is the average of all search queries, search filters, and bots associated with the country for a particular experimental day. Throughout the experimental days, we observed that the misinformation bias scores for SA were consistently higher than the scores for the US. This indicates that, overall, bots in SA routinely received more misinformative SERPs than bots in the US during our study.

\section{Validations and Robustness Analysis}\label{sec:validation-analysis}
In this section, we conducted validation checks to reinforce the robustness of our findings.

\subsection{Validation of Results}

Although our classifier achieves comparable or exceeds the classification performance of models from prior studies (accuracy of 0.85), it has a 15\% error rate, which may affect our findings. In this section, we aimed to confirm the reliability of our findings. To do so, we selected one misinformation topic. We manually annotated the top 10 search results, enabling us to compare the significance test results from fully manually annotated data with those from the classifier-predicted data. We chose the 5G Claims topic because, as shown in Figure \ref{fig:topic_diff}, the topic exhibited statistical significance with the highest effect size (\textit{r}=0.80), indicating that bots in SA received more misinformative search results than their counterparts in the US. For context, Figure \ref{fig:topic_diff} presents the results of the Mann-Whitney U Tests comparing the misinformation bias scores of top-10 search results between the US and SA across each topic.

For the 5G Claims topic, the classifier predicted and labeled 185 videos within the top-10 search results. The first author, as the expert, manually annotated the 185 videos. Subsequently, with complete manual annotations for the 5G Claims topic, we conducted the same analysis, employing the Mann-Whitney U Test to compare the misinformation bias scores for the top-10 search results related to the 5G Claims topic between the US and SA. We obtained a p-value of 6.1e-11 (i.e., p $<$ 0.001) with an effect size \textit{r}=0.84, which is consistent with our findings using classifier-predicted data. This additional analysis provides further evidence of the reliability of our findings despite a 15\% error rate on our classifier.

\begin{table}[t!]
\centering
\small
\begin{tabular}{@{}lccc@{}}
\toprule
\textbf{Condition} & \textbf{\textit{p}} & \textbf{\textit{r}} & \textbf{Mean Rank Diff.}\\ 
\midrule
Main Finding & *** & 0.49 & SA $>$ US \\ 
Case 1 & *** & 0.57 & SA $>$ US \\ 
Case 2 & ** & 0.37 & SA $>$ US \\ 
\bottomrule
\end{tabular}
\caption{Summary of the main finding and validation checks under two alternative treatments of the ``On the COVID-19 origins in Wuhan, China'' label. The results are based on Mann-Whitney U Tests comparing misinformation bias scores between the US and SA for the top-10 videos in the SERPs. Each test reports the p-value (\textit{p}), Mann-Whitney effect size (\textit{r}), and mean rank difference. For example, SA $>$ US indicates that bots in SA received more misinformative SERPs than bots in the US. These validation checks reinforce the robustness of our main findings, showing consistent results regardless of how lab leak theory videos were categorized. Note that: *$P < 0.05$; **$P < 0.01$; ***$P <0.001$.}
\label{tab:validation_checks}
\end{table}

\subsection{Robustness Analysis Under Alternative Treatments of Lab Leak Theory Class}
Since the Lab Leak Theory are often featured heavily as misinformation, we conducted additional validation checks to evaluate the robustness of our main findings under two alternative treatments of the ``On the COVID-19 origins in Wuhan, China'' label. Originally, videos discussing COVID-19 origins were labeled as ``On the COVID-19 origins in Wuhan, China''(see \S \ref{sec:annotation}) and later merged into a neutral ``neither'' class to reflect the lack of consensus about their status as misinformation.\footnote{As of September 2024, \citet{npr_lab_leak} maintained that the Lab Leak Theory is plausible.} Considering the top-10 videos in the SERPs, our main finding showed that bots in SA were exposed to significantly more misinformative SERPs than bots in the US (p $<$ 0.001, effect size \textit{r}=0.49). We replicate this main finding by conducting the same significance test while 1) excluding videos labeled under the ``On the COVID-19 origins in Wuhan, China'' class in the SERPs (Case 1), and 2) treating the class as misinformation (Case 2). Table \ref{tab:validation_checks} summarizes the results of the main finding and validation checks, demonstrating consistent results across all scenarios, regardless of how lab leak theory videos were categorized.

\textbf{Case 1: Eliminating the class.} We removed all videos labeled as ``On the COVID-19 origins in Wuhan, China'' from the dataset and repeated the significance test to compare the misinformation bias scores for the top-10 search results between the US and SA. Our results confirmed that bots in SA still encountered significantly more misinformative SERPs compared to bots in the US (p $<$ 0.001, effect size \textit{r}=0.57), consistent with our main findings.

\textbf{Case 2: Merging the class with misinformation.} We merged the ``On the COVID-19 origins in Wuhan, China'' class with the ``supporting misinformation'' class and conducted the same significance test. Again, SA bots were exposed to significantly more misinformative SERPs compared to bots in the US (p $<$ 0.01, effect size \textit{r}=0.37), consistent with our main findings. 

These validation checks reinforce the robustness of our findings, demonstrating that bots in SA encountered significantly more misinformative SERPs than bots in the US regardless of how lab leak theory videos were categorized.

\begin{table*}[ht!]
\small
\centering
\begin{tabular}{cp{2.5cm}p{7.5cm}p{5cm}}  
\toprule
\textbf{Scale} & \textbf{Annotation Label} & \textbf{Descriptions} & \textbf{Example Video (Title and ID)} \\ 
\midrule
-1 & Opposing COVID-19 Misinformation (Opposing) & 
  The video opposes, debunks, satirizes OR provides countervailing contexts to COVID-19 misinformation OR disseminates/endorses public health authorities' policies OR explains the scientific contexts surrounding the public health policy and the COVID-19 pandemic OR provides countervailing contexts to the usage of misleading terms, such as ``Kungflu'' and ``China Virus''. &   
  1. Fox Allows Lunatic to Imply Kim Jong-Un Created Cronavirus (\textit{Video ID: 8EGVzWRic1I})
  
  2. The Monday Times: Coronavirus Car, Ghosts, Tik Tok, 5G Conspiracy | The Daily Show With Trevor Noah (\textit{Video ID: 2totKKH4770})\\
\midrule
0 & Neutral COVID-19 Information (Neutral) & 
  Videos that broadly cover anything related to the COVID-19 pandemic, but do not support nor oppose COVID-19 misinformation. For example, videos that report on the number of COVID-19 cases, discuss remote work hours during a pandemic, etc. & 
  1. Coronavirus outbreak: Hospitals in China swamped with patients (\textit{Video ID: 9qAFanPeKtE})
  
  2. Africa Unite Against COVID 19 (\textit{Video ID: 6EMEEn65JPg})\\
\midrule
1 & Supporting COVID-19 Misinformation (Supporting) & 
  The video supports OR provides evidence for COVID-19 misinformation OR suggests alternative treatment, prevention, and cure without scientific evidence OR exaggerates a claim to suggest a misinformative narrative OR promotes the usage of misleading terms without countervailing contexts. & 
  1. KT The Arch Degree Break's Down The Coronavirus! (MUST SEE) (\textit{Video ID: ulncPbTgPlo})
  
  2. Origin of the CCP Virus: Fauci Emails Revealed; Ivermectin: A Vaccine Alternative?(\textit{Video ID: edT1HYMjnzA})\\
\midrule
2 & On the COVID-19 origins in Wuhan, China (COVID-19 origins) & 
  Any videos that hypothesize OR cover the origins of COVID-19 in Wuhan, China as the main premise of the video (e.g. natural origins theory, lab leak theory) OR provide evidence for the lab leak theory OR trying to debunk/satirize the lab leak theory. Videos that speculate beyond the origin theories of the virus (e.g., COVID-19 was intentionally engineered as a bioweapon in the Wuhan Lab) are marked as supporting COVID-19 misinformation. & 
  1. Progressive Lab Leak Theory (\textit{Video ID: 0iNZE6BEoXs})
  
  2. Why the US is taking a second look at the 'lab leak' theory about COVID-19 (\textit{Video ID: ZRV\_uYS2Buc}) \\
\midrule
3 & Irrelevant & 
  Any video whose content is not related to the COVID-19 pandemic and vaccines. & 
  1. Dengue: Laboratory-acquired case reported in North Carolina, according to study (\textit{Video ID: PxFJ3YSgG0Y})
  
  2. The Nuremberg Code (\textit{Video ID: Eu\_3Cbf7f8Y}) \\
\midrule
4 & YouTube video in a language other than English (Non-English) & 
  The video's title, description, and/or contents in a language other than English. & 
  1. Can Coronavirus be Biological Weapon? Video Analysis by Major General S B Asthana (Retd) (\textit{Video ID: WiHmN\_DuQhQ})
  
  2. Shiva $\mid$ The Arrogant Kung Fu Fighter $\mid$ Episode 54 $\mid$ Download Voot Kids App (\textit{Video ID: ISfm5pAWmxs})\\
\midrule
5 & URL not accessible & 
  Youtube Video URL is not accessible at the time of annotation. & 
  - \\
\bottomrule
\end{tabular}
\caption{Table containing our 7-point annotation labels. For each annotation label, we provide the scale value, description, and two example videos.}
\label{tab:annotation}
\end{table*}

\begin{table*}
\centering
\small
\resizebox{\textwidth}{!}{%
\begin{tabular}{ccp{6.5cm}p{4.8cm}ccc|ccc}  
\toprule
\textbf{ID} & \textbf{Model} & \textbf{Features} & \textbf{Vectorizer} & \textbf{Zero-/Few-Shot} & \textbf{Metadata} & \textbf{Temp.} & \textbf{Acc.} & \textbf{F1-M} & \textbf{F1-W} \\ 
\midrule
\multicolumn{10}{c}{\textit{SVM with Individual Features (FastText)}} \\
\midrule
0 & SVM & Title & FastText & -- & -- & -- & 0.64 & 0.64 & 0.64 \\
1 & SVM & Description & FastText & -- & -- & -- & 0.60 & 0.60 & 0.60 \\
2 & SVM & Transcript & FastText & -- & -- & -- & 0.58 & 0.58 & 0.58 \\
3 & SVM & Tags & FastText & -- & -- & -- & 0.62 & 0.62 & 0.62 \\
4 & SVM & Comments & FastText & -- & -- & -- & 0.49 & 0.47 & 0.47 \\
\midrule
\multicolumn{10}{c}{\textit{SVM with Feature Combinations (FastText)}} \\
\midrule
5 & SVM & Title, Description & FastText & -- & -- & -- & 0.68 & 0.67 & 0.68 \\
6 & SVM & Title, Transcript & FastText & -- & -- & -- & 0.71 & 0.71 & 0.71 \\
7 & SVM & Title, Tags & FastText & -- & -- & -- & 0.71 & 0.71 & 0.71 \\
8 & SVM & Title, Comments & FastText & -- & -- & -- & 0.66 & 0.66 & 0.66 \\
9 & SVM & Title, Description, Transcript & FastText & -- & -- & -- & 0.69 & 0.69 & 0.69 \\
10 & SVM & Title, Description, Tags & FastText & -- & -- & -- & 0.68 & 0.68 & 0.68 \\
11 & SVM & Title, Description, Comments & FastText & -- & -- & -- & 0.66 & 0.66 & 0.66 \\
12 & SVM & Title, Description, Transcript, Tags & FastText & -- & -- & -- & 0.69 & 0.69 & 0.69 \\
13 & SVM & Title, Description, Transcript, Comments & FastText & -- & -- & -- & 0.67 & 0.67 & 0.67 \\
\midrule
\multicolumn{10}{c}{\textit{SVM with All Features$^*$ (Various Vectorizers)}} \\
\midrule
14 & SVM & All Features$^*$ & Count(unigram) & -- & -- & -- & 0.64 & 0.64 & 0.64 \\
15 & SVM & All Features$^*$ & Count(unigram, bigram) & -- & -- & -- & 0.68 & 0.68 & 0.68 \\
16 & SVM & All Features$^*$ & Count(unigram, bigram, trigram) & -- & -- & -- & 0.68 & 0.69 & 0.69 \\
17 & SVM & All Features$^*$ & TFIDF(unigram) & -- & -- & -- & 0.78 & 0.78 & 0.78 \\
18 & SVM & All Features$^*$ & TFIDF(unigram, bigram) & -- & -- & -- & 0.78 & 0.78 & 0.78 \\
19 & SVM & All Features$^*$ & TFIDF(unigram, bigram, trigram) & -- & -- & -- & 0.77 & 0.76 & 0.77 \\
20 & SVM & All Features$^*$ & Word2Vec & -- & -- & -- & 0.76 & 0.76 & 0.76 \\
21 & SVM & All Features$^*$ & FastText & -- & -- & -- & 0.71 & 0.71 & 0.72 \\
\midrule
\multicolumn{10}{c}{\textit{SVM with Feature Combinations (TFIDF)}} \\
\midrule
22 & SVM & Title, Description & TFIDF(unigram) & -- & -- & -- & 0.70 & 0.70 & 0.70 \\
23 & SVM & Title, Description & TFIDF(unigram, bigram) & -- & -- & -- & 0.72 & 0.72 & 0.72 \\
24 & SVM & Title, Transcript & TFIDF(unigram) & -- & -- & -- & 0.73 & 0.73 & 0.73 \\
25 & SVM & Title, Transcript & TFIDF(unigram, bigram) & -- & -- & -- & 0.75 & 0.75 & 0.75 \\
26 & SVM & Title, Tags & TFIDF(unigram) & -- & -- & -- & 0.70 & 0.70 & 0.70 \\
27 & SVM & Title, Tags & TFIDF(unigram, bigram) & -- & -- & -- & 0.71 & 0.71 & 0.71 \\
28 & SVM & Title, Description, Transcript & TFIDF(unigram) & -- & -- & -- & 0.73 & 0.73 & 0.73 \\
29 & SVM & Title, Description, Transcript & TFIDF(unigram, bigram) & -- & -- & -- & 0.76 & 0.76 & 0.76 \\
30 & SVM & Title, Description, Tags & TFIDF(unigram) & -- & -- & -- & 0.72 & 0.72 & 0.72 \\
31 & SVM & Title, Description, Tags & TFIDF(unigram, bigram) & -- & -- & -- & 0.75 & 0.75 & 0.76 \\
32 & SVM & Title, Description, Transcript, Tags & TFIDF(unigram) & -- & -- & -- & 0.75 & 0.75 & 0.75 \\
33 & SVM & Title, Description, Transcript, Tags & TFIDF(unigram, bigram) & -- & -- & -- & 0.78 & 0.78 & 0.78 \\
34 & SVM & Title, Description, Transcript, Comments & TFIDF(unigram) & -- & -- & -- & 0.75 & 0.75 & 0.75 \\
35 & SVM & Title, Description, Transcript, Comments & TFIDF(unigram, bigram) & -- & -- & -- & 0.76 & 0.76 & 0.76 \\
\midrule
\multicolumn{10}{c}{\textit{XGBoost with Feature Combination (TFIDF)}} \\
\midrule
36 & XGB & Title, Description & TFIDF(unigram) & -- & -- & -- & 0.68 & 0.68 & 0.68 \\
37 & XGB & Title, Description & TFIDF(unigram, bigram) & -- & -- & -- & 0.67 & 0.67 & 0.67 \\
38 & XGB & Title, Transcript & TFIDF(unigram) & -- & -- & -- & 0.71 & 0.71 & 0.71 \\
39 & XGB & Title, Transcript & TFIDF(unigram, bigram) & -- & -- & -- & 0.69 & 0.69 & 0.69 \\
41 & XGB & Title, Tags & TFIDF(unigram) & -- & -- & -- & 0.69 & 0.69 & 0.69 \\
42 & XGB & Title, Tags & TFIDF(unigram, bigram) & -- & -- & -- & 0.69 & 0.69 & 0.69 \\
43 & XGB & Title, Description, Transcript & TFIDF(unigram) & -- & -- & -- & 0.70 & 0.69 & 0.69 \\
44 & XGB & Title, Description, Transcript & TFIDF(unigram, bigram) & -- & -- & -- & 0.73 & 0.73 & 0.73 \\
45 & XGB & Title, Description, Tags & TFIDF(unigram) & -- & -- & -- & 0.74 & 0.74 & 0.74 \\
46 & XGB & Title, Description, Tags & TFIDF(unigram, bigram) & -- & -- & -- & 0.75 & 0.75 & 0.76 \\
47 & XGB & Title, Description, Transcript, Tags & TFIDF(unigram) & -- & -- & -- & 0.71 & 0.71 & 0.71 \\
48 & XGB & Title, Description, Transcript, Tags & TFIDF(unigram, bigram) & -- & -- & -- & 0.72 & 0.72 & 0.72 \\
49 & XGB & Title, Description, Transcript, Comments & TFIDF(unigram) & -- & -- & -- & 0.73 & 0.73 & 0.73 \\
50 & XGB & Title, Description, Transcript, Comments & TFIDF(unigram, bigram) & -- & -- & -- & 0.72 & 0.72 & 0.72 \\
51 & XGB & All Features$^*$ & TFIDF(unigram) & -- & -- & -- & 0.73 & 0.73 & 0.73 \\
52 & XGB & All Features$^*$ & TFIDF(unigram, bigram) & -- & -- & -- & 0.73 & 0.73 & 0.73 \\
\midrule
\multicolumn{10}{c}{\textit{DeBERTa Models}} \\
\midrule
53 & DeBERTa (base) & Title, Description, Transcript, Tags & -- & -- & -- & -- & 0.81 & 0.81 & 0.81 \\
54 & DeBERTa (large) & Title, Description, Transcript, Tags & -- & -- & -- & -- & \textbf{0.85} & \textbf{0.85} & \textbf{0.85} \\
\midrule
\multicolumn{10}{c}{\textit{GPT-4 Turbo Models}} \\
\midrule
55 & GPT-4 Turbo & Title, Description, Transcript, Tags & -- & Zero-Shot & Complete & 0 & 0.76 & 0.76 & 0.76 \\
56 & GPT-4 Turbo & Title, Description, Transcript, Tags & -- & Zero-Shot & Truncated & 0 & 0.78 & 0.78 & 0.78 \\
57 & GPT-4 Turbo & Title, Description, Transcript, Tags & -- & Zero-Shot & Truncated & 0.2 & 0.79 & 0.79 & 0.79 \\
58 & GPT-4 Turbo & Title, Description, Transcript, Tags & -- & Zero-Shot & Truncated & 0.7 & 0.77 & 0.77 & 0.77 \\
59 & GPT-4 Turbo & Title, Description, Transcript, Tags & -- & Few-Shot & Truncated & 0 & 0.78 & 0.78 & 0.78 \\
60 & GPT-4 Turbo & Title, Description, Transcript, Tags & -- & Few-Shot & Truncated & 0.2 & 0.78 & 0.78 & 0.78 \\
61 & GPT-4 Turbo & Title, Description, Transcript, Tags & -- & Few-Shot & Truncated & 0.7 & 0.79 & 0.79 & 0.79 \\
\bottomrule
\end{tabular}}
\caption{Performances on our held-out test set (273 videos) across Support Vector Machines (SVM), XGBoost (XGB), DeBERTa-v3, and GPT-4 Turbo (Version 1106) on various features, vectorizers, and parameters. We trained and evaluated 62 models. For GPT-4, we experimented with providing complete vs. truncated video metadata (first 500 tokens of each video metadata) in the prompts; these are indicated by the values ``Truncated'' and ``Complete'' in the column ``Metadata.'' The highest performance value is bolded; DeBERTa-v3-large (Index 54) performed the best. Note that Temp: Temperature, Acc.: Accuracy, F1-M: Macro F1-score, F1-W: Weighted F1-score, and All Features$^*$:  Title, Description, Transcript, Tags, Comments.}
\label{tab:classifier-full-results}
\end{table*}

\begin{table*}[!ht]
    \centering
    \begin{tabular}{p{0.9\textwidth}}
    \toprule 
    \emph{\textbf{Zero-shot prompt}} \\
    You are a public health expert with comprehensive knowledge of the COVID-19 pandemic and the misinformation surrounding it. You are aware of the various health misinformation in videos on YouTube. You are tasked with carefully analyzing the provided video metadata to discern whether the YouTube video falls into one of three labels: opposing the COVID-19 misinformation with scientific information and public health guidance, supporting the COVID-19 misinformation, or presenting neutral information. \\\\
    Using the provided label descriptions and examples, please evaluate the YOUTUBE VIDEO METADATA and assign a label. Below, we provided bulleted descriptions and a few examples for each label, defining what videos would fall into the label category:\\\\

    OPPOSES COVID-19 MISINFORMATION: [OPPOSING DESCRIPTION]\\\\

    SUPPORTS COVID-19 MISINFORMATION: [SUPPORTING DESCRIPTION]\\\\

    NEUTRAL INFORMATION:[NEUTRAL DESCRIPTION]\\\\
    
    Now, given what you learned from the label descriptions and examples above, please evaluate and assign a label to the YOUTUBE VIDEO METADATA and provide justification on your label with direct excerpts(s) from the YOUTUBE VIDEO METADATA. FORMAT your response as a JSON object in the following structure [(LABEL, EXCERPTS, JUSTIFICATION)].\\\\

    YOUTUBE VIDEO METADATA starts here *****:\\
    Video Title: [TITLE]\\
    Video Description: [DESCRIPTION]\\
    Video Transcript: [TRANSCRIPT]\\
    Video Tags: [TAGS]\\
    \bottomrule
    \end{tabular}
    \caption{Zero-shot prompt to determine whether a YouTube video opposes COVID-19 misinformation, supports COVID-19 misinformation, or is neutral information. For each label, we employed the same descriptions provided to AMT workers on their annotation task. Note that ``Neutral Information'' here refers to the merged category, consisting of ``Neutral COVID-19 Information,'' ``COVID-19 origins,'' and ``Irrelvant'' labels. Please see subsection ``Consolidating from 5-classes to 3-classes'' in \S \ref{sec:classifier} for more information.}
    \label{tab:zero-shot-prompt}
\end{table*}

\begin{table*}[!ht]
    \centering
    \begin{tabular}{p{0.9\textwidth}}
    \toprule 
    \emph{\textbf{Few-shot prompt}} \\
    You are a public health expert with comprehensive knowledge of the COVID-19 pandemic and the misinformation surrounding it. You are aware of the various health misinformation in videos on YouTube. You are tasked with carefully analyzing the provided video metadata to discern whether the YouTube video falls into one of three labels: opposing the COVID-19 misinformation with scientific information and public health guidance, supporting the COVID-19 misinformation, or presenting neutral information. \\\\
    Using the provided label descriptions and examples, please evaluate the YOUTUBE VIDEO METADATA and assign a label. Below, we provided bulleted descriptions and a few examples for each label, defining what videos would fall into the label category:\\\\

    OPPOSES COVID-19 MISINFORMATION: [OPPOSING DESCRIPTION]\\\\

    SUPPORTS COVID-19 MISINFORMATION: [SUPPORTING DESCRIPTION]\\\\

    NEUTRAL INFORMATION:[NEUTRAL DESCRIPTION]\\\\

    We provide five examples of the task, each featuring video metadata, label, and reasoning.\\
    EXAMPLE 1 starts here ****:\\
    VIDEO\_TITLE: [EXAMPLE1\_VIDEO\_TITLE]\\
    VIDEO\_DESCRIPTION: [EXAMPLE1\_VIDEO\_DESCRIPTION]\\
    VIDEO\_TRANSCRIPT: [EXAMPLE1\_TRANSCRIPT]\\
    VIDEO\_TAGS: [EXAMPLE1\_TAGS]\\
    LABEL: [EXAMPLE1\_LABEL]\\
    REASONING: [EXAMPLE1\_REASONING]\\
    ...\\
    ...\\
    ...\\
    EXAMPLE 5 starts here ****:\\
    VIDEO\_TITLE: [EXAMPLE5\_VIDEO\_TITLE]\\
    VIDEO\_DESCRIPTION: [EXAMPLE5\_VIDEO\_DESCRIPTION]\\
    VIDEO\_TRANSCRIPT: [EXAMPLE5\_TRANSCRIPT]\\
    VIDEO\_TAGS: [EXAMPLE5\_TAGS]\\
    LABEL: [EXAMPLE5\_LABEL]\\
    REASONING: [EXAMPLE5\_REASONING]\\\\
    
    Now, given what you learned from the label descriptions and examples of the task, please evaluate and assign a label to the YOUTUBE VIDEO METADATA and provide justification on your label with direct quote(s) from the YOUTUBE VIDEO METADATA. FORMAT your response as a JSON object in the following structure [(LABEL, QUOTE, JUSTIFICATION)].\\\\

    YOUTUBE VIDEO METADATA starts here *****:\\
    Video Title: [TITLE]\\
    Video Description: [DESCRIPTION]\\
    Video Transcript: [TRANSCRIPT]\\
    Video Tags: [TAGS]\\
    \bottomrule
    \end{tabular}
    \caption{Few-shot prompt to determine whether a YouTube video opposes COVID-19 misinformation, supports COVID-19 misinformation, or is neutral information. For each label, we employed the same descriptions and examples provided to AMT workers on their annotation task. Note that ``Neutral Information'' here refers to the merged category, consisting of ``Neutral COVID-19 Information,'' ``COVID-19 origins,'' and ``Irrelvant'' labels. Please see subsection ``Consolidating from 5-classes to 3-classes'' in \S \ref{sec:classifier} for more information.}
    \label{tab:few-shot-prompt}
\end{table*}

\end{document}